\documentclass[journal=jacsat,manuscript=article,layout=twocolumn]{achemso}

\usepackage{amssymb}
\usepackage{graphicx}
\usepackage{dcolumn}
\usepackage[version=3]{mhchem}
\usepackage{textcomp, gensymb}

\author{Christian P. N. Tanner}
\affiliation{Department of Chemistry, University of California, Berkeley, CA 94720, USA}

\author{James K. Utterback}
\affiliation{Department of Chemistry, University of California, Berkeley, CA 94720, USA}
\alsoaffiliation{Present Address: Sorbonne Université, CNRS, Institut des NanoSciences de Paris, INSP, 75005 Paris, France}

\author{Joshua Portner}
\affiliation{Department of Chemistry, James Franck Institute, and Pritzker School of Molecular Engineering, University of Chicago, Chicago, IL 60637, USA}

\author{Igor Coropceanu}
\affiliation{Department of Chemistry, James Franck Institute, and Pritzker School of Molecular Engineering, University of Chicago, Chicago, IL 60637, USA}

\author{Avishek Das}
\affiliation{Department of Chemistry, University of California, Berkeley, CA 94720, USA}
\alsoaffiliation{Present Address: AMOLF, Science Park 104, 1098 XG, Amsterdam, The Netherlands}

\author{Christopher J. Tassone}
\affiliation{Stanford Synchrotron Radiation Lightsource, SLAC National Accelerator Laboratory, Menlo Park, CA 94025, USA}

\author{Samuel W. Teitelbaum}
\affiliation{Department of Physics, Arizona State University, Tempe, AZ 85287, USA}

\author{David T. Limmer}
\affiliation{Department of Chemistry, University of California, Berkeley, CA 94720, USA}
\alsoaffiliation{Chemical Sciences Division, Lawrence Berkeley National Laboratory, Berkeley, CA 94720, USA}
\alsoaffiliation{Materials Sciences Division, Lawrence Berkeley National Laboratory, Berkeley, CA 94720, USA}
\alsoaffiliation{Kavli Energy NanoSciences Institute, University of California, Berkeley, CA 94720, USA}

\author{Dmitri V. Talapin}
\affiliation{Department of Chemistry, James Franck Institute, and Pritzker School of Molecular Engineering, University of Chicago, Chicago, IL 60637, USA}
\alsoaffiliation{Center for Nanoscale Materials, Argonne National Laboratory, Argonne, IL 60517, USA}

\author{Naomi S. Ginsberg}
\email{nsginsberg@berkeley.edu}
\affiliation{Department of Chemistry, University of California, Berkeley, CA 94720, USA}
\alsoaffiliation{Department of Physics, University of California, Berkeley, CA 94720, USA}
\alsoaffiliation{Molecular Biophysics and Integrated Bioimaging Division, Lawrence Berkeley National Laboratory, Berkeley, CA 94720, USA}
\alsoaffiliation{Materials Sciences and Chemical Sciences Division, Lawrence Berkeley National Laboratory, Berkeley, CA 94720, USA}
\alsoaffiliation{Kavli Energy NanoSciences Institute, University of California, Berkeley, CA 94720, USA}
\alsoaffiliation{STROBE, NSF Science \& Technology Center, Berkeley, CA 94720, USA}

\title{\textit{In situ}  X-ray scattering reveals coarsening rates of superlattices self-assembled from electrostatically stabilized metal nanocrystals depend non-monotonically on driving force}
\begin{document}


\begin{abstract}

Self-assembly of colloidal nanocrystals (NCs) into superlattices (SLs) is an appealing strategy to design hierarchically organized materials with promising 
functionalities. Mechanistic studies are still needed to uncover the design principles for SL self-assembly, but such studies have been difficult to perform due to the fast time- and short length scales of NC systems. To address this challenge, we developed an apparatus to directly measure the evolving phases \textit{in situ} and in real time of an electrostatically stabilized Au NC solution before, during, and after it is quenched to form SLs using small angle X-ray scattering (SAXS). By developing a quantitative model, we fit the time-dependent scattering patterns to obtain the phase diagram of the system and the kinetics of the colloidal and SL phases as a function of varying quench conditions. The extracted phase diagram is consistent with particles whose interactions are short in range relative to their diameter. We find the degree of SL order is primarily determined by fast (sub-second) initial nucleation and growth kinetics, while coarsening at later times depends non-monotonically on the driving force for self-assembly. We validate these results by direct comparison with simulations and use them to suggest dynamic design principles to optimize crystallinity within a finite time window. The combination of this measurement methodology, quantitative analysis, and simulation should be generalizable to elucidate and better control the microscopic self-assembly pathways of a wide range of bottom-up assembled systems and architectures. 

\end{abstract}

\subsubsection{Keywords:}
soft condensed matter, nanocrystals, self-assembly, coarsening, X-ray scattering, \textit{in situ} measurement

\maketitle
\section*{Introduction}

Colloidal nanocrystal (NC) building blocks can be used to self-assemble a variety of functional, ordered structures or superlattices (SLs)\cite{boles_self-assembly_2016,murray_self-organization_1995,shevchenko_structural_2006,smith_self-assembled_2009,bian_shape-anisotropy_2011,wang_colloids_2012,fan_self-assembled_2010,wang_self-assembled_2012,murray_synthesis_2000,santos_macroscopic_2021,van_blaaderen_template-directed_1997} with potential energy and optoelectronic applications, such as solar cells\cite{gur_air-stable_2005}, sensors\cite{wang_materials_2018}, catalysts\cite{arandiyan_pt_2015}, and displays\cite{zhu_assembly_2017,begley_bridging_2019}. In order to reliably control and direct the self-assembly of SLs, a detailed understanding of the microscopic interactions between NCs and of the thermodynamic and kinetic landscapes of the self-assembly process is necessary. Electrostatic forces are important in the self-assembly of a variety of SL structures, but describing the interactions between charged NCs in electrolytic solutions remains a challenge. The interactions between micron scale charged colloids in electrolytic solutions are well understood within Derjaguin Landau Verwey Overbeek (DLVO) theory, which describes the interactions as a linear combination of van der Waals attraction, electrostatic repulsion mediated by the electrolytic solution, and steric repulsion\cite{israelachvili_intermolecular_2011}. DLVO theory, however, has limited applicability to nanoscale systems\cite{silvera_batista_nonadditivity_2015}, especially at high ion concentrations, since the ions in solution are finite in size relative to the NCs and can no longer be considered point charges. As a result, it is challenging to describe and predict the phase behavior of charged NCs in electrolytic solutions as well as the kinetics of their self-assembly into SLs. Therefore, experimental studies of NC SL phase coexistence and the kinetics associated with particular pathways through the associated phase diagram are needed.

Measuring NC systems is challenging due to their fast time- and short length scales. Typically, the phase behavior of colloidal systems is measured using optical techniques\cite{pusey_phase_1986,gasser_real-space_2001,calderon_experimental_1993,monovoukas_experimental_1989,dinsmore_phase_1995,verhaegh_fluid-fluid_1996}, but these approaches are ineffective for NCs that fall below the diffraction limit. Synchrotron X-ray scattering is in principle an attractive method to address these challenges since it provides nanoscale structural information down to ms time scales\cite{weidman_kinetics_2016,graceffa_sub-millisecond_2013,karim_synthesis_2015,wu_tuning_2018}. Yet, few \textit{in situ} X-ray scattering studies of NC SL self-assembly exist due to the difficult sample geometries required to follow the full self-assembly process. For electrostatically induced self-assembly, these difficulties include simultaneously processing the initial colloidal suspension, mixing it with reagents, and protecting it from air and humidity, all the while probing a homogenized volume in a thin, X-ray compatible chamber over the full course of the transformation. Furthermore, the few existing studies focus on NCs with organic surface ligands that self-assemble into SLs \textit{via} spin coating\cite{weidman_kinetics_2016}, solvent evaporation\cite{korgel_small-angle_1999,weidman_kinetics_2016,lu_resolving_2012,lokteva_real-time_2021,josten_superlattice_2017,smilgies_superlattice_2017,geuchies_situ_2016}, or growth from solution\cite{wu_high-temperature_2017,abecassis_gold_2008,marino_temperature-controlled_2023,grote_x-ray_2021,qiao_situ_2023}, but not on the effect of electrostatic forces on self-assembly. In addition, previous studies primarily measured the kinetics of SL self-assembly under specific conditions and could not also obtain the associated phase diagrams, limiting their ability to directly correlate the kinetics with phase diagram features. 

Here, we nevertheless use small angle X-ray scattering (SAXS) with an apparatus we developed to measure \textit{in situ} and in real time solutions of electrostatically stabilized colloidal Au NCs before, during, and after they are quenched to varying degrees to form SLs. Using a model that we developed for multicomponent solution scattering, we fit the time-dependent SAXS patterns to quantitate the relative amounts of the colloidal and SL phases as well as the crystal structure and crystallinity of the SLs. The SL product and remaining colloidal NC fractions enable mapping of the system phase diagram as a function of colloid concentration and quench depth, which provides insight into the effective range and depth of the interparticle interactions in this system. In addition, the combination of our apparatus, self-assembly protocol, and data analysis techniques enable us to determine how the kinetics of the self-assembly process at different quench depths impact the resulting SL crystallinity and yield and how to alter the protocol for optimal outcomes. Brownian dynamics simulations corroborate the experiments and help to reveal the underlying interparticle interactions and resulting mechanisms of SL growth and annealing. This work presents a generalizable strategy to more completely elucidate the microscopic interactions and self-assembly pathways of a wide range of NC SLs, enabling the design of 
structures with improved optical, electronic, and mechanical functionalities.

\section*{Results and Discussion}

\begin{figure*}
\includegraphics[width=16cm]{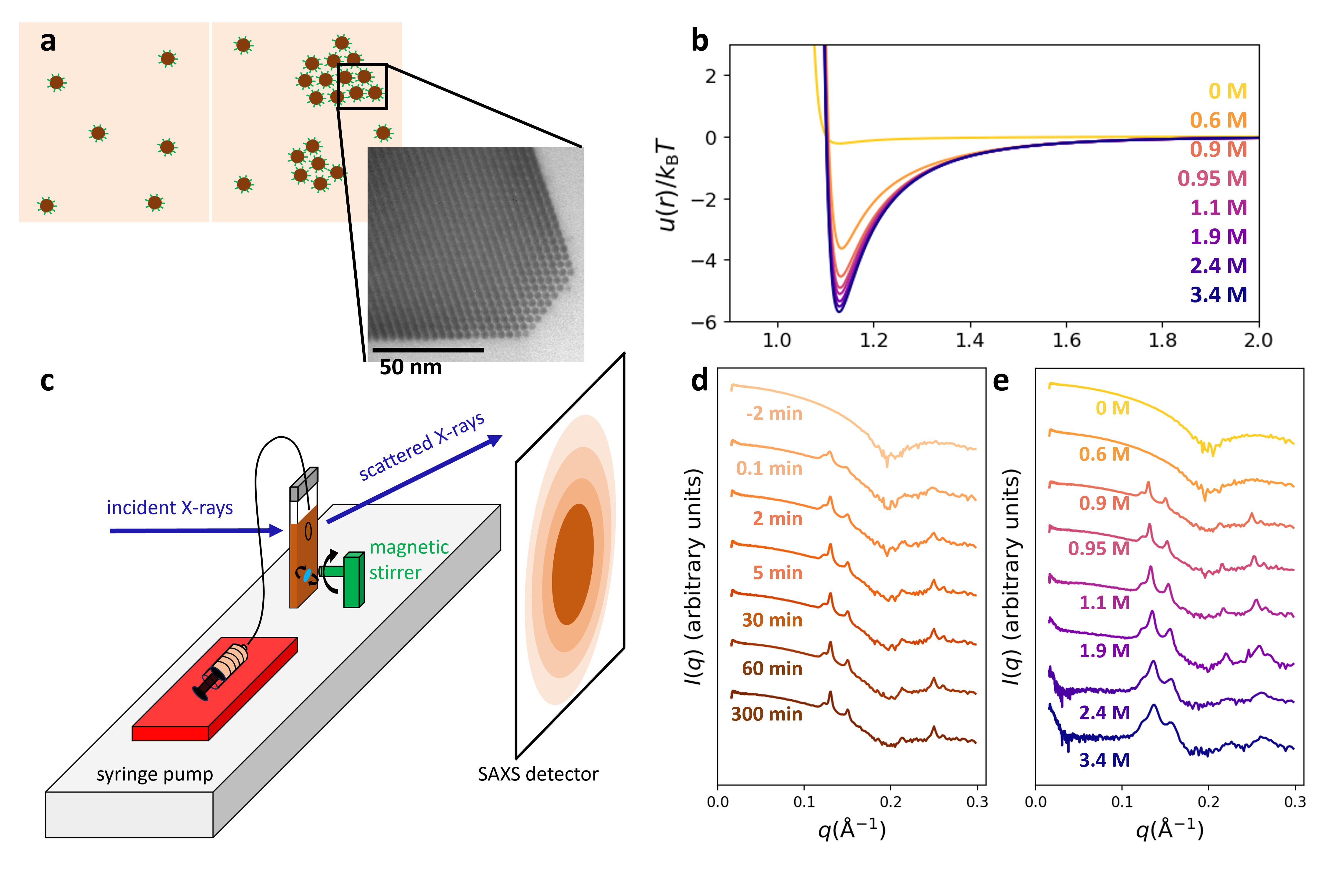}
\caption{\textit{In situ}  monitoring of electrostatic self-assembly of NC SLs. (a) Schematic of colloidal NCs (brown) with thiostannate surface ligands (green lines) in hydrazine (tan) (left) and coexistence of colloidal NCs with SLs (right). SL domains are not drawn to scale. Inset: Transmission electron microscope image of Au NC SL. Reprinted in part with permission from Reference 41. Copyright 2022 The American Association for the Advancement of Science. (b) DLVO interaction potentials vs. relevant solution ionic strengths. Here, $u_0$ is the minimum value of a given curve. (c) Schematic of gas-tight apparatus for measuring X-ray scattering of self-assembly \textit{in situ}. (d) SAXS patterns as a function of time during a typical experiment at a final solution ionic strength of 0.90 M and NC volume fraction of 0.004. (e) SAXS patterns at long times post-quench from \textit{in situ} experiments performed at NC volume fractions $\sim$0.002 and at different solution ionic strengths.}
\end{figure*}

To determine the effect of electrostatics on the self-assembly of SLs, we study Au NCs with thiostannate ($\mathrm{Sn_2S_6^{4-}}$) surface ligands colloidally suspended in hydrazine ($\mathrm{N_2H_4}$), a polar solvent with a dielectric constant of 52 at room temperature (\textbf{Figure 1a} left). Unlike typical NCs with organic surface ligands, these NCs have charged ligands and NC-NC interactions are thus controlled \textit{via} electrostatic forces.\cite{kovalenko_colloidal_2009,coropceanu_self-assembly_2022} In order to quench the system to generate a condensed phase (\textbf{Figure 1a} right), we add additional $\mathrm{(N_2H_5)_4Sn_2S_6}$ salt solution to the initial $\sim$50 mg/mL NC suspension, which screens the electrostatic repulsion between NCs and creates overall attractive interactions between NCs. In this study, we systematically varied the final ionic strength of the solution, $I=\frac{1}{2}\sum_{n=1}^{N}c_nz_n^2$, from 0.6 mol/L (M) to 3.4 M. Here, $c_n$ is the concentration of ion species $n$ in molar, $z_n$ is the valency of ion $n$, and $N$ is the number of different ion species in solution. The ionic strength controls the quench depth, \textit{i.e.}, the driving force for self-assembly, which, formally, is the potential energy difference between an Au NC in the SL and colloidal phases. 

Despite its limitations (see Supporting Information) DLVO theory still offers a qualitative understanding of the interparticle interactions as a function of quench depth. Within its framework, the interparticle interactions are determined by linear combinations of the van der Waals attraction, steric repulsion, and electrostatic repulsion between NCs. While the strength of attraction between NCs is set by the van der Waals force, the electrostatics modulate the potential by adding a tunable repulsive force between NCs. For the range of solution ionic strengths studied here, DLVO theory predicts short-range interaction potentials with well depths $u_0 \sim 3-6$ $k_\mathrm{B}T$ (\textbf{Figure 1b}, see \textbf{Methods} for details of the calculations). The thiostannate ligands, which are a key ingredient enabling the self-assembly of ordered SLs, contribute to the high NC surface charge density (\textit{i.e.}, the magnitude of the electrostatic repulsion term) as well as the effective size of the NCs (the steric repulsion term) (see Supporting Information).

To experimentally monitor the self-assembly of Au NC SLs, we developed an apparatus for use with synchrotron small angle X-ray scattering (SAXS). The gas-tight apparatus consists of a quartz cuvette, with thinned, X-ray transparent windows, connected to a syringe on a syringe pump \textit{via} tubing inserted into a septum (\textbf{Figures 1c}, \textbf{S1a}, \textbf{S1b}). A magnetic stirrer rotates a stir bar in the cuvette to homogenize the solution and prevent SLs from sinking to the bottom (\textbf{Figure S1c}). In a typical experiment, the cuvette is initially filled with Au NCs colloidally dispersed in hydrazine, and we quench the system by using the syringe pump to inject over $\sim5-12$ seconds a controlled amount of $\mathrm{(N_2H_5)_4Sn_2S_6}$ salt dissolved in hydrazine. We collect two-dimensional (2D) SAXS detector images before, during, and after the quench at a rate of one image every 5 seconds for up to two hours post-quench. By azimuthally averaging the 2D SAXS detector images, we obtain one-dimensional SAXS patterns, $I(q)$, that describe the scattered intensity as a function of the scattered X-ray momentum transfer, $q$. The time-dependent series of SAXS patterns provide an ensemble average measure of the evolving phase coexistence of the system over the course of SL self-assembly as well as detailed information about each phase, such as the size distribution of the NCs and crystal structure of the SLs. We next describe the observed time- and quench-dependent SAXS patterns and then share how they were fit to a quantitative model to extract the phase diagram of the system and the kinetics of SL formation.

Typical one-dimensional SAXS patterns as a function of time post-quench are shown in \textbf{Figure 1d}. Before the quench, the SAXS patterns show the scattering from the 4.5 nm-diameter colloidal NCs. Immediately after quenching the system by bringing the solution ionic strength up to 0.9 M and within the $\sim$s time resolution of the experiment, \textit{fcc} SL Bragg peaks emerge at the expense of the colloidal phase. The SL Bragg peaks continuously grow and narrow over the entire measurement window. \textbf{Figure 1e} shows $\sim$1 hour post-quench ($\sim$equilibrium) SAXS patterns from 6 experiments at a series of final solution ionic strengths. At a final solution ionic strength of 0.6 M, the system remained purely colloidal. As the quench depth (\textit{i.e.}, solution ionic strength) increases, the widths of the near-equilibrium SL Bragg peaks increase and their peak positions shift to higher $q$, indicating the SLs are smaller, more disordered, and have smaller lattice constants\cite{warren_x-ray_1990}.

In order to extract information from the time-dependent SAXS patterns, we developed and used a model to quantitatively fit them. Specifically, we model the background-subtracted scattered intensity as  $I(q) = I_{\mathrm{colloid}}(q) + I_{\mathrm{SL}}(q)$, where $I_{\mathrm{colloid}}(q)$ is the scattered intensity from dilute, polydisperse hard spheres and $I_{\mathrm{SL}}(q)$ is the scattered intensity from finite-sized \textit{fcc} crystals (See \textbf{Methods}, \textbf{Figures S2}, \textbf{S3}). This model fits the data well at all time points and quench depths (see \textbf{Figure S4}), as shown for a selection of final solution ionic strengths in \textbf{Figure 2a}. This fitting scheme allows us to extract the relative amount of NCs in the colloidal and SL phases as well as the degree of crystallinity of the SL phase as a function of time and quench depth. 

\begin{figure*}
\includegraphics[width=18cm]{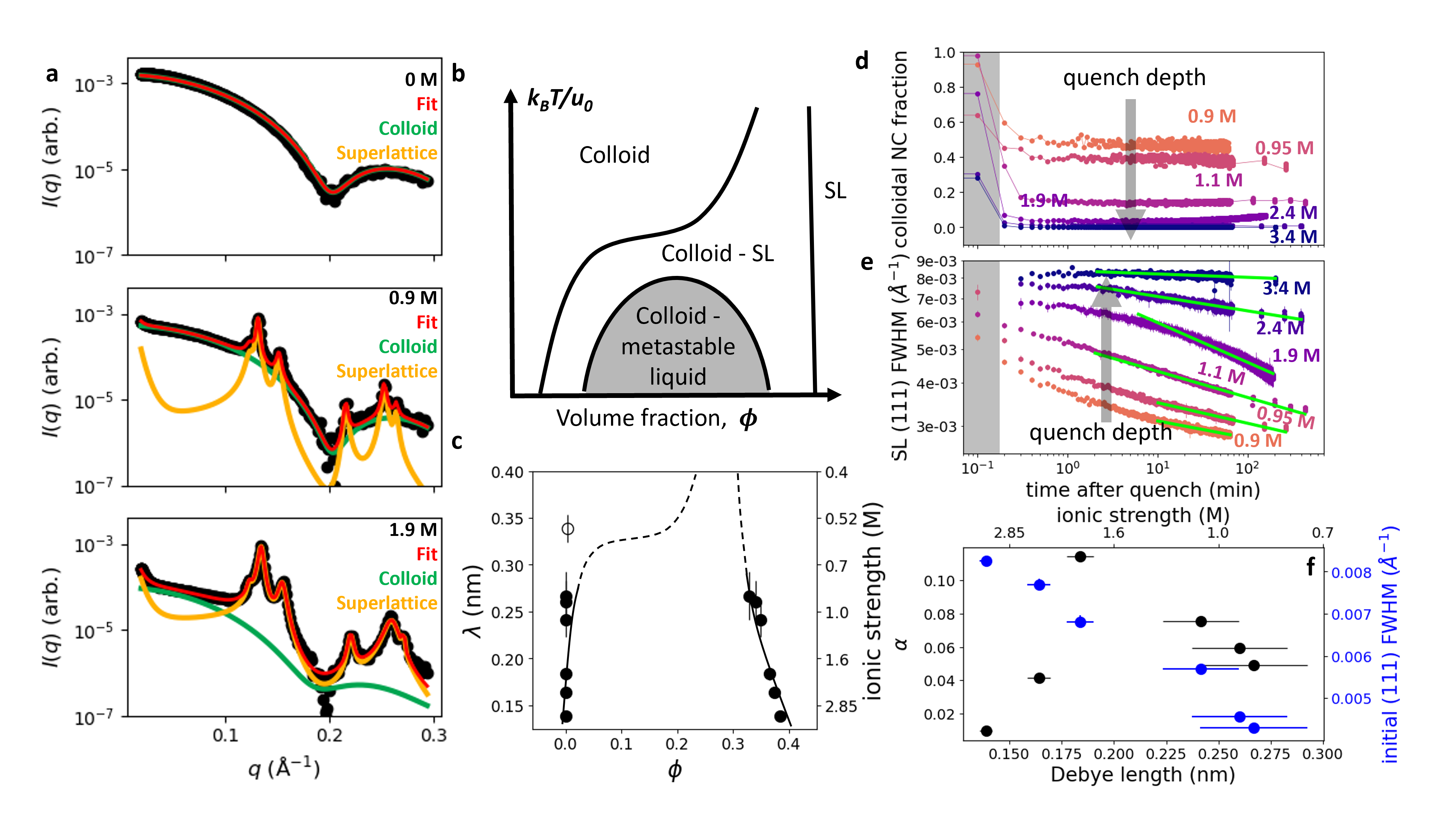}
\caption{Quantitative analysis of time- and quench-dependent SAXS patterns. (a) Quantitative fits of model to colloidal NC and SL SAXS patterns at solution ionic strengths of 0, 0.9, and 1.9 M. (b) Schematic phase diagram for spherical particles interacting \textit{via} short-range potentials. (c) Quantitative phase diagram for electrostatically stabilized Au NCs obtained from experimental observations (black data points). The open circle indicates a phase diagram location where the system is purely colloidal. Vertical error bars indicate the standard deviations of the Debye lengths, $\lambda$, of the solutions based on the uncertainty of the volume and concentration of the injected salt solution. Horizontal error bars indicating the standard deviations of the colloidal and SL volume fractions due to the same volume uncertainty and uncertainty from SAXS fitting are smaller than the size of the black data points. Black phase boundary curves are sketched as a visual guide based on the location of the black data points. The dashed continuation of these phase boundary curves are sketched to aid comparison to \textbf{Figure 2b}. (d) Fraction of NCs remaining in the colloidal phase as a function of time for a series of quench depths ranging from 0.9 M (orange) to 3.4 M (purple). Grey region indicates the period during which salt injection took place. The same series is treated in panels (e) and (f). (e) FWHM of the SL (111) Bragg peak as a function of time post-quench. Vertical error bars indicate the standard deviations of the FWHM from SAXS pattern fitting uncertainty. Green lines are power law fits to the FWHM at late times. Grey region same as in (d). (f) FWHM power law exponents, $\alpha$, as a function of Debye length (black) and the FWHM of SLs upon completion of salt injection (blue). Black vertical error bars indicating the standard deviation in $\alpha$ due to uncertainty from fitting and blue vertical error bars indicating standard deviation of the FWHM from SAXS pattern fitting uncertainty are both smaller than the size of the data points. Black and blue horizontal error bars are the same as in (c).}
\end{figure*}

A cartoon of the expected phase diagram for monodisperse spherical particles with interparticle potentials short in range relative to their diameter is shown in \textbf{Figure 2b}\cite{ten_wolde_enhancement_1997}. The phase diagram consists of binodal (phase boundary) curves that specify the presence and density of each phase as a function of the volume fraction of particles in solution, $\phi$, and the effective temperature, $k_\mathrm{B}T/u_0$, where $u_0$ is the depth of the interparticle potential. The ratio $u_0/k_\mathrm{B}T$ formally defines the quench depth. For example, at high $k_\mathrm{B}T/u_0$ and low $\phi$, the colloidal phase is the only stable phase. As $k_\mathrm{B}T/u_0$ is lowered or $\phi$ is increased and the left-most binodal is crossed, the solid, or SL, phase becomes thermodynamically stable and exists in equilibrium with the colloidal (gas-like) phase. In addition to the colloidal and SL phases, if an additional binodal is crossed (grey in \textbf{Figure 2b}), then a liquid phase, which consists of densely packed yet fluid colloidal particles with no long range order, exists as well. Unlike in phase diagrams for typical atomic systems where the interactions are long-range relative to the size of the atom, the colloid-metastable liquid binodal in the phase diagram in \textbf{Figure 2b} is situated below the colloid-SL binodal. Consequently, the liquid phase is not thermodynamically stable, but previous simulation and experimental work has shown it can exist metastably and even act as a precursor to SL formation\cite{ten_wolde_enhancement_1997,wedekind_optimization_2015,haxton_crystallization_2015,karthika_review_2016,savage_experimental_2009,zhang_charge-controlled_2012,pagan_phase_2005,vekilov_two-step_2010,lee_entropic_2019}. The exact location of the colloid-metastable liquid binodal relative to the colloid-SL binodal depends primarily on the range of the interparticle potential\cite{noro_extended_2000}. Specifically, as the range of interaction decreases, the metastable liquid phase becomes less stable, and the colloid-metastable liquid binodal peaks at lower $k_\mathrm{B}T/u_0$.

To determine the thermodynamic landscape of the self-assembly process studied here, we calculated, from our experimental data and quantitative fitting, the points on the binodal curves that constitute the phase diagram of the Au NC system in \textbf{Figure 2c}. These points specify the volume fraction, or density, of NCs in the colloidal or SL phases, respectively, as a function of quench depth. We use the Debye length $\lambda=\sqrt{\frac{\epsilon_r \epsilon_0 k_\mathrm{B}T}{\sum_{n=1}^{N}c_nz_n^2e^2}}$ of the solution as a figure of merit for the quench depth on the vertical axis since it combines information on the (varying) ionic strength of the solution and the dielectric constant of the solvent.  Here, $\epsilon_r$ is the solvent dielectric constant, $\epsilon_0$ is the vacuum permittivity, $e$ is the charge of an electron, and $c_n$ and $z_n$ are the same as in the equation for the solution ionic strength. Although this choice does not incorporate the impact of the steric and van der Waals contributions from the NCs into $k_\mathrm{B}T/u_0$, these forces should not vary from quench to quench since the same NC stock solution was used for all measurements. To determine the points on the low density side of the phase diagram in \textbf{Figure 2c}, we compare the scattered X-ray intensity vs. $q$ of the colloidal phase after equilibrating $\sim$ 1 hour post-quench to the scattered colloidal intensity prior to the quench. From this ratio we obtain the fraction of NCs remaining in the colloidal phase as a function of the quench depth. Multiplying this value by the total volume fraction of NCs in the system, we obtain the volume fraction of the colloidal phase at equilibrium with the SL phase for each quench depth. These volume fractions provide the horizontal axis values of the left-hand colloid-SL binodal in \textbf{Figure 2c}. For a given quench depth, the volume fraction of the SL phase determines the points on the high density colloid-SL binodal, which we obtain from the position of the \textit{fcc} (111) Bragg peak, $q_{111}$, using $\phi=4V/a^3$, where $V$ is the volume of a NC excluding ligands and $a$ is the lattice constant of the SL ($a = 2\pi\sqrt{3}/q_{111})$. The low density colloid-SL binodal rises steeply with $\phi$ at low $\phi$ from $\phi \sim 0.0$ to $0.0009$. On the high density side, the colloid-SL binodal rises less steeply from $\phi \sim 0.38$ to $0.33$. In order to resolve the colloid-SL binodals between $\phi \sim 0.01$ and $0.3$, larger total NC concentrations than studied here would be required. We did not do so in this work due to greater difficulty stabilizing the colloidal phase at high NC concentrations. Nevertheless, our observations do constrain the extent of the rise of the low-density binodal at greater $\phi$ than shown with our data points based on the fact that a final solution ionic strength above $\sim$ 0.6 M ($\lambda < 0.34$ nm) is needed to generate SLs (open circle in \textbf{Figure 2c} and see in more detail in \textbf{Figure S5}). We sketched the curves between $\phi$ $\sim$0.01 and $\sim$0.3 with dashing as an interpolation between the solid curves dictated by experimental data; the dashed curves are not meant to be quantitative. At all time points and quench conditions studied here, we did not observe a liquid phase in the SAXS patterns, and as a result our measured phase diagram consists solely of colloid-SL binodals.

To characterize the time evolution of the colloidal and SL phases, we also extracted the kinetics of each phase. The kinetics of the colloidal phase, which are anti-correlated with those of the SL phase (not shown), in \textbf{Figure 2d} are extracted similarly to the method described above to determine the equilibrium fraction of colloidal NCs for the phase diagram in \textbf{Figure 2c}. We find that the fraction of NCs remaining in the colloidal phase decreases monotonically with the quench depth. While the fraction of NCs remaining in the colloidal phase as a function of time in any given quench also decreases monotonically, the decrease is very small in magnitude following the salt injection period shaded in gray in \textbf{Figure 2d}. This finding suggests that the colloid fractions following the initial quench approach the thermodynamically expected values.

In \textbf{Figure 2e}, we determined the full width at half maximum (FWHM) of the \textit{fcc} (111) SL Bragg peak as a function of time and quench depth. This quantity encodes both the coherence length of the SL, \textit{i.e.}, the typical length scale of a crystalline domain, and the degree of crystallinity of those domains. We focused our analysis to the SL (111) peak since the limited \textit{q}-resolution and signal to noise of the higher order Bragg peaks limited the reliability of more involved analysis methods such as Williamson-Hall and Debye-Waller analyses. The SL (111) FWHM immediately after the injection increase monotonically with the quench depth (\textbf{Figure 2f} blue points) and decrease monotonically with time at all studied quench depths (orange to purple respectively from 0.9 to 3.4 M in \textbf{Figure 2e}). We fit the late-time behavior to a power law, $\mathrm{FWHM} \sim t^{-\alpha}$  (\textbf{Figure 2e} green lines, see Supporting Information), and find that $\alpha$ depends non-monotonically on the quench depth (\textbf{Figure 2f} black points).
At the shallowest SL-producing quench to $\lambda = 0.267$ nm (0.9 M),  $\alpha = 0.049 \pm 0.001$. As the quench depth increases, $\alpha$ increases and reaches a maximum of $0.110 \pm 0.001$ at $\lambda = 0.184$ nm (1.9 M). As the quench depth continues to increase, $\alpha$ decreases and has a value of $0.010 \pm 0.001$ at the deepest quench depth at $\lambda = 0.139$ nm (3.4 M). 

To determine the relationships between the SL (111) FWHM, the colloidal NC fractions, and the underlying self-assembly mechanisms, we simulated the self-assembly process at two different quench depths. The simulations were performed in the NVT ensemble, representing the NCs as spherical particles interacting through a coarse-grained short-range attractive Morse potential. Starting from a homogeneous phase, we quenched the NCs to two different depths of the interaction potential ($u_0/k_\mathrm{B}T$) and studied the dynamics of the growth of dense and ordered clusters (see \textbf{Methods} and \textbf{SI} for details of the simulations). Snapshots of the simulations are shown in \textbf{Figures 3a,b}. For the shallow quench simulation, we induced nucleation with a spherical \textit{fcc} seed, while in the deeper quench nucleation of spherical liquid droplets occurred spontaneously. SLs nucleated from within the liquid droplets and subsequently grew and annealed. Every 20 Brownian time units, we calculated the expected X-ray scattering patterns of the system and fit them using our model (see \textbf{Methods} and \textbf{Figure S6}). The simulated colloidal NC fractions for both quenches in \textbf{Figure 3c} decrease monotonically, indicating that NCs are transferring from the colloidal phase to the SL phase as the SLs grow. \textbf{Figure 3d} shows the simulated \textit{fcc} (111) Bragg peak FWHM for the two quench depths. We uncover two kinetic regimes that we classify with power laws. As indicated by the dotted green lines, the $\sim$sub-ms kinetics follow a $t^{-\alpha}$ power law with $\alpha = 0.435 \pm 0.013$ and $\alpha = 0.447 \pm 0.008$ for the shallow and deeper quenches, respectively. We do not observe this regime in the experiments due to the limited time resolution associated with finite-time injection and the SAXS detector acquisition rate. At longer $\sim$ms times, the power laws of the simulated kinetics become more similar to the experimental values obtained on $\sim$s to hour time scales with $\alpha = 0.116 \pm 0.002$ for the shallow quench simulation and $\alpha = 0.063 \pm 0.003$ for the deeper quench simulation, each indicated with solid green lines in \textbf{Figure 3d}.

With access to the real space positions of every simulated NC, we calculate the ensemble-averaged crystallinity of NCs in the SL phase by tracking for each NC a Steinhardt-Nelson order parameter\cite{steinhardt_bond-orientational_1983} for orientations of bonds with neighboring NCs (see \textbf{Methods}). Unlike the FWHM of the Bragg peaks, the average crystallinity decouples the finite size of the SLs from the degree of order of the SL domains. The average crystallinities of SLs in the two simulations are shown in \textbf{Figure 3e}. We find that the SLs in the deeper quench simulation have lower average crystallinity than the SL in the shallow quench simulation. As indicated by the green lines, the late time average crystallinity kinetics also follow a power law with $\alpha = 0.014 \pm 0.001$ for the shallow quench and $\alpha = 0.035 \pm 0.002$ for the deeper quench. 

\begin{figure*}
\includegraphics[width=6cm]{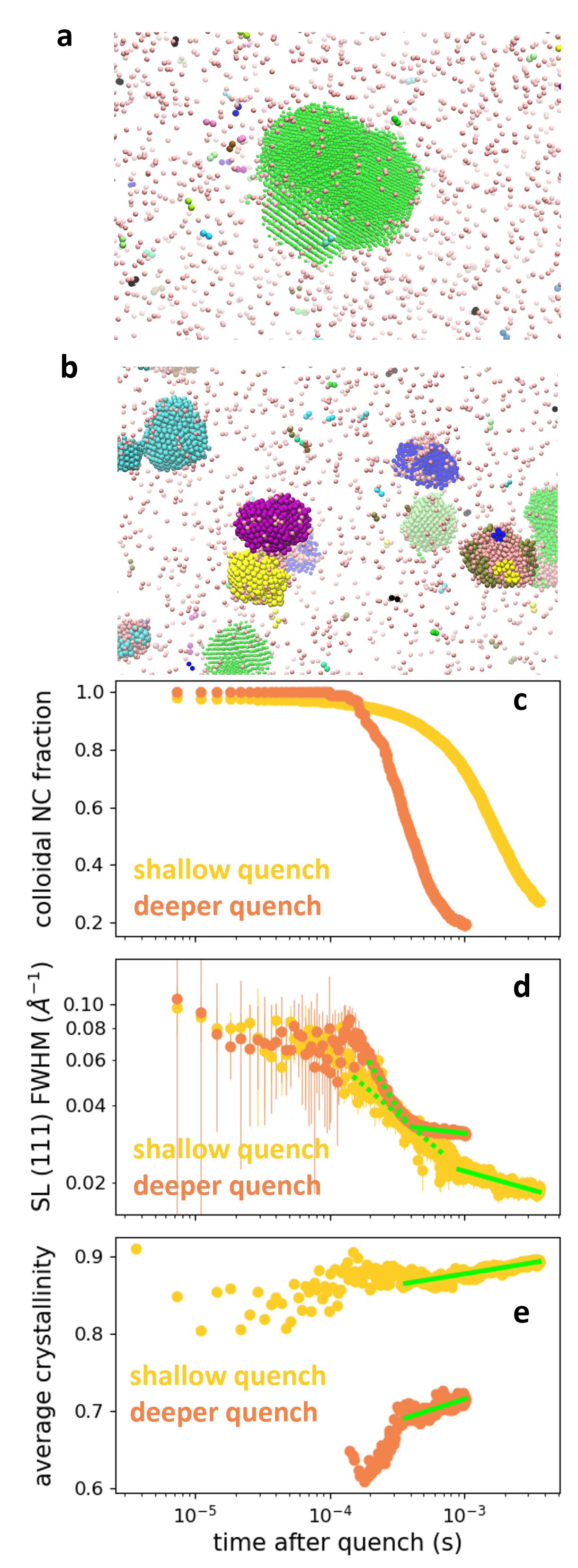}
\caption{Simulations of SL self-assembly. (a) Snapshot of the shallow quench simulation. The different colors correspond to different crystalline clusters (see \textbf{Methods} for cluster determination) with colloidal NCs shown in light pink. The visualized NC size is reduced by 70\% from 4.5 nm to aid visualization. (b) Snapshot of the deeper quench simulation at same scale as in (a). Colors and NC sizes determined as in (a). (c) Simulated colloidal NC fractions vs.\,time for shallow quench (yellow) and deeper quench (orange). (d) Simulated SL \textit{fcc} (111) Bragg peak FWHM vs.\,time for shallow quench (yellow) and deeper quench (orange). Vertical error bars indicate standard deviation in SL (111) FWHM due to uncertainty from fitting procedure. Power law fits to early times are shown in dotted green and to later times are shown in solid green.  (e) Average crystallinity of particles in the condensed phase vs.\,time for both quenches. Colors are the same as in (c) and (d). Power law fits to late times are shown in solid green.}
\end{figure*}


Having described our observations, SAXS analysis strategies, and corroborating simulations, we turn toward a discussion of our findings and their implication on NC SL self-assembly. We begin by commenting on the features of the phase diagram extracted from the experiments as well as the insight the phase diagram provides into the effective range and strength of NC-NC interparticle interactions in this system. We discuss these insights in relation to the predictions from DLVO theory. Next, we summarize the different kinetic regimes (nucleation/growth and coarsening) of SL self-assembly and how they impact the resulting SLs. Finally, by comparing how the early- and late-time kinetics vary as a function of the quench depth in experiment and in simulation, we propose kinetic design principles for optimal SL self-assembly in finite time. 

With our measurement and analysis protocol, we are able to directly and quantitatively reveal a substantial portion of the phase diagram for the electrostatically stabilized Au NCs studied in this work. The phase diagram is consistent with the expectation that as the quench depth increases, the equilibrium volume fraction of the colloidal phase decreases as more colloidal NCs are incorporated into SLs (\textbf{Figures 2c, S5}). In addition, the volume fraction of the SL phase increases as the quench depth increases. This trend could be due to the depth ($u_0$) of the interparticle potentials becoming greater at deeper quenches or due to NC size-selectivity during self-assembly, \textit{i.e.}, larger NCs condensing before smaller ones due to stronger van der Waals attraction. We propose that NC size-selectivity is the most likely reason based on estimation using a statistical analysis of NC diameters, nearest neighbor distances between NCs in the SL phase at different quench depths, and scalings of van der Waals attraction vs. NC diameter (see \textbf{Figure S7} and associated text). One additional limitation of obtaining the phase diagram with the methods described in this work is that the systems $\sim$1-2 hours post-quench at different quench depths may not all be similarly close to equilibrium. Although the colloidal NC fractions following the initial quench approach the thermodynamically expected values (\textbf{Figure 2d}), if the systems measured were not exactly at equilibrium, the true low-density colloid-SL binodals would be located at slightly smaller $\phi$ than we obtained. While it is difficult to know exactly at which densities the equilibrium states will be, we estimate that the colloidal NC fractions at equilibrium are within a few percent of the corresponding fractions $\sim$1 hour post quench (see Supporting Information). At deeper quenches, the SL (111) FWHM increase, indicating that the SLs are further away from their equilibrium structures. As a result, the high-density colloid-SL binodal we extracted may very slightly underestimate the true equilibrium SL density. 

Despite the small uncertainty in the precise locations of the phase diagram binodals, we can use the phase diagram and the evolving phase coexistence to obtain insight into the nature of the interparticle interactions governing the self-assembly process. For example, we find no presence of a liquid phase in any of the SAXS patterns obtained after equilibration (or at any time) at any quench depth. The absence of a thermodynamically stable liquid phase implies that these NCs interact \textit{via} short-range attractive potentials. Even though we do not experimentally observe a metastable liquid phase, the phase diagram for this system is consistent with phase diagrams of particles with short-range interactions because it consists solely of regions where either the colloidal phase is the only stable phase or where colloid and SL coexist. The phase diagram is inconsistent with those of hard spheres or particles interacting \textit{via} long-range interactions because SLs form at volume fractions lower than $\phi = 0.49$, the freezing density of hard spheres, and there is no region of colloid-liquid coexistence at shallow quenches as there would be for particles with long-range attractive interactions. One reason we may not observe the metastable liquid phase is that it may convert entirely into SLs beneath the time resolution of the measurement ($\sim$5 s). Indeed, in the deeper quench simulation, SLs nucleate and grow from within liquid droplets on average in $<$ 200 $\mu$s (\textbf{Figure S8a}). Another possibility is that the range of the interparticle potential is sufficiently short that it suppresses the colloid-metastable liquid binodal cartooned in \textbf{Figure 2b} far enough below the colloid-SL binodal that it is experimentally inaccessible under our quenching conditions. In other words, our quenches may only place the system in the colloid-SL coexistence region between the colloid-SL and colloid-metastable liquid binodals in \textbf{Figure 2b} (see \textbf{Figure S8b}). Our previous work on the self-assembly of SLs from these NCs showed the formation of dense agglomerations of NCs after similar amounts of salt solution as used in this study were added to the initial colloidal NC suspensions in combination with a MeCN anti-solvent.\cite{coropceanu_self-assembly_2022} The additional MeCN may have quenched the solution to low enough $\lambda$ to access the metastable liquid binodal. While that more aggressive quench protocol followed its own specific kinetic trajectory, it is also possible that the liquid state could be kinetically forbidden using the less aggressive protocols in this present work at even higher ionic strengths, if quenches too shallow to cross the metastable liquid-colloid binodal were already to lead to kinetic arrest.

In principle, knowledge of the exact location of the colloid-metastable liquid binodal relative to the colloid-SL binodal enables determination of the effective range of the interparticle potential. Since we did not observe any liquid phase and therefore could not map out a colloid-metastable liquid binodal, we are unable to precisely specify the range of the interparticle potential. Nevertheless, since we do not observe a thermodynamically stable liquid phase, we can infer from the Noro-Frenkel law of corresponding states\cite{noro_extended_2000} that the center-to-center range of the interparticle potential must be $<$ $1.2\sigma_{\mathrm{eff}}$, where $\sigma_{\mathrm{eff}}$ is the effective size of a NC, including its ligand shell. Combined with the measured nearest neighbor distances between NCs in the SL phase (\textbf{Figure S7b}), we estimate the effective size of the NCs to be $\sim5.6-5.9$ nm, resulting in an effective center-to-center range of interparticle interactions of $\sim6.7-7.1$ nm, \textit{i.e.}, no greater than 1.2$\sigma_{\mathrm{eff}}$. Interestingly, as a result, the DLVO theory predictions of the range of the interactions for this system under the quench conditions studied here (\textbf{Figure 1b}) were reasonable. DLVO theory, however, overestimates the depth ($u_0$) of the interparticle potentials. We base this conclusion on the following reasoning. First, DLVO predicts a well depth of $\sim3k_\mathrm{B}T$ for a solution ionic strength of 0.6 M (\textbf{Figure 1b}). At 0.6 M, however, the system remained purely colloidal (\textbf{Figure 1e} orange curve), which could only occur if the well depth were less than $\sim2k_\mathrm{B}T$\cite{klotsa_predicting_2011}. In addition, well depths of 2.5 and 2.8 $k_\mathrm{B}T$ used in the simulations respectively resulted in colloidal NC fractions of $\sim$0.27 and 0.19 at the end of the simulated trajectories. These fractions are similar in magnitude to those obtained from experiment $\sim$1 hour after a shallow to intermediate quench. This similarity indicates that smaller values of $u_0$ than those predicted by DLVO theory produce colloidal NC fractions that are consistent with the experimental results (see Supporting Information). These findings suggest that while DLVO theory provides qualitatively accurate predictions for the range of the interparticle potential, it fails quantitatively in its prediction of the depths of the interparticle potentials for this charged NC system. Although DLVO predictions may agree with the inferred potential depths of the deeper experimental quenches (perhaps as deep as 6 $k_\mathrm{B}T$), based on our simulations and additional computational work\cite{haxton_crystallization_2015,klotsa_predicting_2011}, we expect the potential depths of shallower quenches to be closer to 2.5 $k_\mathrm{B}T$, which explicitly disagrees with DLVO. The inferred interparticle potentials between NCs following shallow, intermediate, and deep quenches in experiments are shown in \textbf{Figures 4a, S9}. \textit{In situ} experimental approaches combined with quantitative analysis and simulation tools as presented in this work thus provide a means by which to determine the nature of the underlying interactions. 

While the long-time behavior of the system post-quench provides insight into the nature of the interparticle interactions and general thermodynamic landscape of the system as a function of $\lambda$ and $\phi$, the full time evolution of the self-assembly process reveals two distinct kinetic regimes. On sub-ms time scales in the simulations, the SL (111) FWHM decrease as power laws with power law exponents $\sim$0.45 corresponding to the initial nucleation and growth of the SLs from the colloidal phase, which we do not resolve experimentally. On $\sim$ms time scales in the simulations and on min-hr time scales in the experiments, the SL (111) FWHM decrease at much smaller power law rates. These slower kinetics correspond to the coarsening stage of the self-assembly process. During this stage, the SL (111) FWHM in the experiments and simulations decrease due to annealing of defects. While coherent X-ray scattering experiments can be used to more explicitly specify the nature of the disorder being annealed,\cite{hurley_situ_2024} this observation could be due to either the average crystallinity increasing within each SL domain or the annealing of grain boundaries separating distinct SL grains within a single polycrystalline SL. The simulated SL (111) FWHM additionally decrease due to SL growth \textit{via} incorporation of NCs from the colloidal phase. These time scales indicate that the SLs in the simulation leave the initial nucleation and growth regime and enter the coarsening regime within a few ms post-quench. This finding also suggests that the SLs observed in the experiments have already entered the coarsening regime during the injection period. 

\begin{figure*}
\includegraphics[width=16cm]{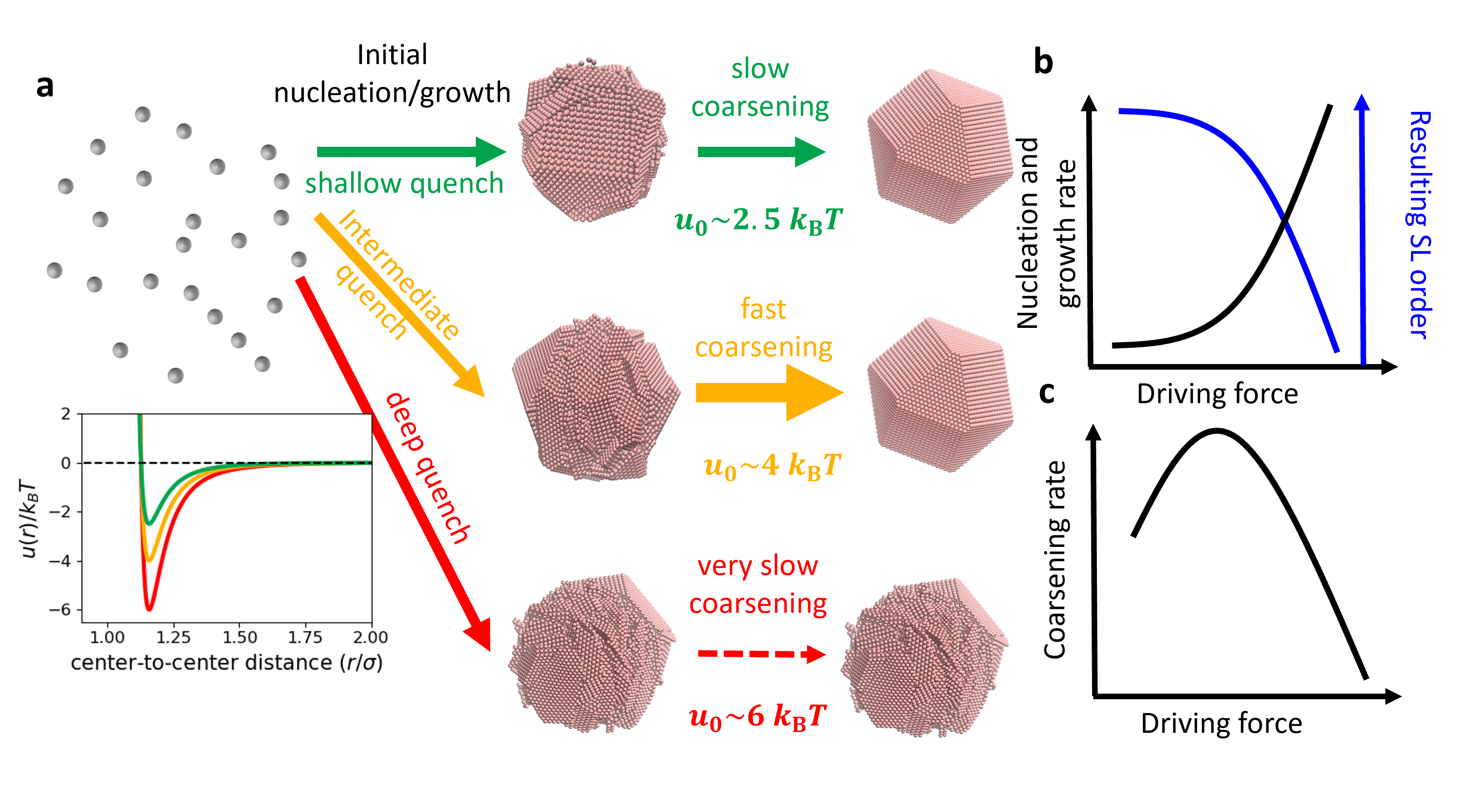}
\caption{SL formation mechanisms and time scales. (a) Top: at shallow quench depths, highly ordered SLs nucleate and grow from solution at time scales $<$ 1 s. These SLs grow and anneal defects slowly over the course of minutes to hours. Middle: at intermediate quench depths, slightly more disordered SLs nucleate initially, and over longer time scales the SLs grow and anneal away the disorder at a faster rate than at shallow quenches. Bottom: in deep quenches, disordered SLs form and are unable to anneal away defects due to kinetic trapping. Inset: interparticle potentials for shallow (green), intermediate (gold), and deep (red) quenches. (b) Sketch of SL nucleation and growth rate and order of the resulting SLs as a function of the driving force for self-assembly. (c) Sketch of rate at which SLs coarsen as a function of the driving force for self-assembly. }
\end{figure*}

In order to further determine the impact of the two kinetic regimes on the self-assembly process, we discuss the kinetic trends as a function of the quench depth. The trends in the initial SL (111) FWHM and the late-time power law exponents as a function of the quench depth in the experiments (\textbf{Figure 2f}) reveal that at shallow quench depths, the initially formed SLs are very ordered and only slowly coarsen over the course of minutes to hours (\textbf{Figure 4a} top). As the quench depth increases, the SLs are initially more disordered but are able to anneal these defects faster than at shallow quenches (\textbf{Figure 4a} middle). As the quench depth increases further, the SLs become much more disordered and can no longer anneal away defects due to kinetic trapping even though the thermodynamic driving force for self-assembly is even stronger (\textbf{Figure 4a} bottom). The simulations show qualitatively similar trends: after the initial nucleation and growth period, the SLs in the simulated deeper quench have larger FWHM (\textbf{Figure 3d}) and lower crystallinity (\textbf{Figure 3e}) than in the shallow quench. Although the simulations cannot access the same long time scales as in the experiments, the SLs in the simulation already enter the coarsening regime within the simulated time frame ($\sim$a few ms). During the coarsening stage in the simulations, the SLs continue to grow \textit{via} incorporation of colloidal NCs and to anneal defects. As a result, the SL (111) FWHM in the shallow quench simulation decrease with a larger power law exponent ($\alpha = 0.116$) than those in the deeper quench simulation ($\alpha = 0.063$) primarily due to faster growth. This behavior is distinct from the behavior on min-hour time scales in experiment since the colloidal NC fractions have plateaued by then, and the decrease in the SL (111) FWHM is primarily due to defect annealing. In fact, during the coarsening stage the simulated deeper quenched SLs anneal various defects and increase their average crystallinity at a faster power law rate ($\alpha = 0.035$) than the SL in the simulated shallow quench ($\alpha = 0.014$). While polydispersity could also play a role in this latter outcome,\cite{schope_effect_2007,auer_prediction_2001} this trend supports the idea that the SL (111) FWHM at intermediate quenches in the experiment decrease at faster rates than those at shallow quenches in experiment due to an increased ability to anneal remaining defects and increase their crystallinity. Although not explicitly simulated here, even deeper quenches would likely lead to kinetic arrest,\cite{haxton_crystallization_2015,klotsa_predicting_2011} which qualitatively agrees with the trend observed experimentally in \textbf{Figures 2e,f} because it would corroborate the non-monotonic behavior in the coarsening rate vs. quench depth.

By combining the kinetic trends as a function of quench depth in the experiment and simulation, we propose emergent design principles for NC SL self-assembly even beyond the electrostatically stabilized system studied here. Typically, the nucleation and growth rate increases exponentially with the quench depth, or driving force for self-assembly (\textbf{Figure 4b} black curve). The conventional wisdom is that larger driving forces lead to faster kinetics, which ultimately results in SLs that are less ordered than SLs assembled more slowly (\textbf{Figure 4b} blue curve). Immediately post-quench, the FWHM of the SLs as a function of quench depth in the experiments (\textbf{Figures 2e,f}) and in the simulations (\textbf{Figure 3d}) support this trend. Surprisingly, the coarsening rate for this system (\textbf{Figure 2f}) does not follow this trend and instead depends non-monotonically on the driving force for self-assembly (\textbf{Figure 4c}). This finding suggests a \textit{kinetic} strategy to improve SL self-assembly. Specifically, using a small driving force to nucleate and grow the initial SLs and then increasing the driving force to better promote \textit{in situ} coarsening should facilitate self-assembly of more highly ordered SLs within a finite amount of time. 

\section*{Conclusion}

In summary, we presented an improved method by which to measure the self-assembly of NC SLs \textit{in situ} and in real-time using synchrotron X-ray scattering. We developed a quantitative model to fit time-dependent SAXS patterns of NCs in the colloidal and SL phases, which enabled us to extract the phase diagram and kinetics of the transformation under different conditions. By combining the insights from simulation with experiment, we have shown the ability to elucidate the effective range and depth of interactions between charged NCs in electrolytic solutions, which are consistent with short ranged potentials. We also found the SL self-assembly kinetics have two regimes (nucleation/growth and coarsening) and that the coarsening kinetics depend non-monotonically on the driving force for self-assembly. Consequently, we propose kinetic strategies to promote ordered SL self-assembly by increasing the driving force for self-assembly as a function of time. 

Identifying this proposed design protocol was only possible thanks to the powerful combination of \textit{in situ} measurement, quantitative analysis, and simulation used in this work, which unveiled the equilibrium properties and non-equilibrium effects that underlie NC SL self-assembly. 
In particular, this approach should be able to reveal the impact of a metastable liquid phase on SL formation and protein crystallization in systems where the metastable liquid either has a longer lifetime or is stable at lower Debye lengths. 
More generally, a similar approach could be used to design protocols for other related systems, such as in protein crystallization\cite{galkin_control_2000,zhang_charge-controlled_2012} and the formation of other hierarchical materials such as metal-\cite{seoane_multi-scale_2016,millange_time-resolved_2010,cheetham_thermodynamic_2018} and covalent-organic frameworks\cite{feng_covalent_2012,li_nucleation_2017,smith_mechanistic_2014}, by determining first how to tune the self-assembly driving force from an understanding of the interparticle/intermolecular interactions and resulting thermodynamic landscape and, second, how to apply the driving force based on the kinetics.
The approach should also be generalizable to elucidate the microscopic pathways and design principles of a variety of nanoscale self-assembly phenomena, for example, in enhancing the complexity of DNA based nanomachines\cite{song_smart_2013} and optimizing nanostructures for drug delivery\cite{shin_covid-19_2020}. 

\section*{Methods}

\textbf{DLVO calculations.} The DLVO interaction potentials in \textbf{Figure 1b} are the sum of three terms: $u(r) = u_{\mathrm{es}}(r) + u_{\mathrm{st}}(r) + u_{\mathrm{vdW}}(r)$, where $u_{\mathrm{es}}(r)$ describes the contribution from the electrostatic interaction between two charged NCs, $u_{\mathrm{st}}(r)$ describes the contribution from the steric overlap of the NCs at small $r$, $u_{\mathrm{vdW}}(r)$ describes the van der Waals attraction between the NCs, and $r$ is the center-to-center distance between two NCs. The electrostatic contribution, $u_{\mathrm{es}}(r) = \frac{64\pi k_\mathrm{B}T R (\sum_{n=1}^{N}c_n)  \tilde{\gamma}^2}{\kappa^2}e^{-\kappa (r-2R)}$, where $R$ is the radius of the NCs, $c_n$ is the concentration of ion $n$, $N$ is the number of different ion species in solution, $\kappa=\lambda^{-1}$, and $\tilde{\gamma}=\tanh{\frac{ze\psi}{4k_\mathrm{B}T}}$ where $z$ is the valency of the NC surface ligand, $e$ is the charge of an electron, and $\psi$ is the NC surface potential. For the calculations in this work, values of $\psi$ such that $\tilde{\gamma}=1$ were used. The van der Waals contribution $u_{\mathrm{vdW}}(r)=-\frac{A}{6}\big(\frac{4R^2}{r^2}+\ln(\frac{r^2-4R^2}{r^2}) \big)$, where $A=1.6$ eV is the Hamaker constant of gold (see Supporting Information).\cite{biggs_measurement_1994} For $u_{\mathrm{st}}(r)$, we use an exponential with a steep cutoff at 0.5 nm (roughly the size of a single ligand molecule): $u_{\mathrm{st}}(r)=ae^{-b(r-\sigma)}$, where $a=5\times10^4$ $k_\mathrm{B}T$ and $b=5.4\times10^{-11}$ nm$^{-1}$.

\textbf{\textit{In situ} SAXS experiments.} All SAXS data were collected at the Stanford Synchrotron Radiation Lightsource (SSRL) at beamline 1-5 with a photon energy of 15 keV and beam size of 600$\times$600 $\mu$m (see Figure S1 for further information). Stock solutions of 4.5 nm Au NCs with $\mathrm{(N_2H_5)_4Sn_2S_6}$ ligands in hydrazine and 0.5 M $\mathrm{(N_2H_5)_4Sn_2S_6}$ salt in hydrazine were prepared following a procedure previously outlined\cite{kovalenko_colloidal_2009,coropceanu_self-assembly_2022}. In a nitrogen filled glovebox, 400-500 $\mu$Ls of a 50 mg/mL ($\phi \sim$0.0026) solution of 4.5 nm Au NCs with $\mathrm{(N_2H_5)_4Sn_2S_6}$ ligands in hydrazine was loaded into a 2 mm path length quartz cuvette with custom 200 $\mu$m thick windows. A small stir bar was placed into the cuvette in the plane of the cuvette and the cuvette was then sealed using a rubber septum and parafilm. A syringe preloaded with a solution of 0.5 M $\mathrm{(N_2H_5)_4Sn_2S_6}$ in hydrazine was attached to the cuvette \textit{via} Teflon tubing through the septum. The tubing-septum interface was sealed with epoxy. The gas-tight apparatus was carefully moved into the beam path and the syringe placed onto a New Era syringe pump (model NE-1000). X-ray scattering data were collected continuously while stirring the solution using a magnetic stirrer from Ultrafast Systems. All X-ray scattering patterns were collected using 1 s exposures at a rate of one pattern every 5 s. For each \textit{in situ}  experiment, after about 5 min of data acquisition, the excess salt in hydrazine solution was injected using the syringe pump at a rate of 847.6 $\mu$L/s. The injection took $\sim$5 - 12 s depending on how much salt was added. The total volume fraction of NCs in solution post-injection varied from $\sim$0.0017 to $\sim$0.0024 depending on the volume of the salt solution that was injected. The apparatus was kept at room temperature (see Supporting Information for additional temperature considerations). Data were continuously acquired after injection for up to two hours. SAXS patterns of cuvettes filled with hydrazine and varying amounts of $\mathrm{(N_2H_5)_4Sn_2S_6}$ salt were taken for background subtraction (see \textbf{Figure S2}). Because the scattering from NCs depends only on the magnitude of the scattered X-ray momentum transfer, $|\vec{q}|=q$, and because the scattering from SLs results from many SLs at different orientations with respect to the X-ray beam, we azimuthally average the 2D SAXS detector images without loss of information to obtain one-dimensional SAXS patterns, $I(q)$, that describe the scattered intensity as a function of $q$. 

\textbf{Modeling of SAXS patterns.} We model the background-subtracted scattered intensity as $I(q) = I_{\mathrm{colloid}}(q) + I_{\mathrm{SL}}(q)$, where $I_{\mathrm{colloid}}(q)$ is the scattered intensity from colloidal NCs and $I_\mathrm{SL}(q)$ is the scattered intensity from finite-sized \textit{fcc} SLs. For $I_\mathrm{colloid}(q)$, we use the form factor for dilute, polydisperse, hard spheres with a Gaussian size distribution. We calculate the form factor using xrsdkit (https://github.com/scattering-central/xrsdkit). For the SL term, we multiply the form factor by the structure factor for a finite-sized \textit{fcc} SL. We model the SL structure factor as the sum of a set of Lorentzian line shapes each centered on a respective Bragg peak of an \textit{fcc} lattice and an additional $q^{-4}$ term. We fit our model to the experimental SAXS patterns, $I(q)$, to obtain the SL Bragg peak positions and FWHM and the relative amount of NCs in the colloidal and SL phases.  For more information and justification of this model to describe the scattering from finite-sized SLs, see Supporting Information and \textbf{Figure S3} and associated text.

\textbf{Simulations of self-assembly.} Simulations of SL growth and annealing were performed with an underdamped Langevin dynamics in the NVT ensemble in a cubic periodic box using the LAMMPS software\cite{thompson_lammps_2022}. NCs were represented as 10976 spherical particles with pairwise volume exclusion interactions given by a Weeks-Chandler-Andersen (WCA) potential\cite{weeks_role_1971}. Additionally, NCs interact also \textit{via} a pairwise attractive short-range Morse potential. At a given temperature $T$, the diffusive timescale for NCs is given by  $\tau=\gamma\sigma^2/k_\mathrm{B}T$ where $\sigma$ is the NC diameter, $\gamma$ is the friction coefficient in Langevin dynamics, and $k_\mathrm{B}$ is Boltzmann constant. To compare to experimental timescales, we assume that the NCs follow Stokes’ law of diffusion, with friction coefficient relating to the solvent viscosity $\eta$ as $\gamma=3\pi\eta\sigma$. We then use $\sigma = 4.5$ nm, $\eta= 0.876 \times 10^{-3}$ Pa-s, and $T = 300$ K to obtain $\tau=0.18$ $\mu$s. NCs were initially equilibrated in the gas phase before being adiabatically quenched to  $u_0=2.5k_\mathrm{B}T$ and $u_0=2.8k_\mathrm{B}T$ for the shallow and deeper quench, respectively. In the case of the shallow quench, we used a spherical defect-free \textit{fcc} crystal of size 200 NCs as a seed to start crystal growth and annealing. For more details on the simulations, see Supporting Information.

\textbf{Crystallinity calculation.} We tracked crystalline order during the simulated self-assembly trajectories by computing for each NC its local bond-orientational order, $\psi_6^{(i)}$,  using Steinhard-Nelson order parameters\cite{steinhardt_bond-orientational_1983}. At each time frame, we define a NC’s nearest neighbors as all other NCs within a center-to-center cutoff distance of $1.5\sigma$. A NC was defined to be locally crystalline if either its $\psi_6^{(i)}$ parameter was above a cutoff of 0.7\cite{steinhardt_bond-orientational_1983}, or if its neighbor was locally crystalline. All locally crystalline particles are then classified into clusters of direct or indirect neighbors. All reported results about the crystallinity of NCs in the SL phase are computed only over NCs in these dense locally ordered clusters containing at least 100 NCs. For more details on the local bond-orientational order calculations, see Supporting Information.

\textbf{Simulated scattering patterns.} We calculated the structure factor, $S(q)$, of the particles in the simulations every $20\tau$ time units using the formula $S(\mathbf{q}) = \frac{1}{N}|\sum_{n=1}^{N}e^{-i\mathbf{q}\cdot \mathbf{r}_n}|^2$, where $\mathbf{r}_n$ is the location of particle $n$ and $N=10976$. We then average over shells of constant $q = |\mathbf{q}|$ to obtain $S(q)$. We fit our model for the SL structure factor to the simulated $S(q)$ to extract the FWHM and Bragg peak positions (see \textbf{Figure S6}).

\section*{Conflicts of Interest}
There are no conflicts to declare.

\section*{Acknowledgments}

We thank A. Liebman-Pel\'{a}ez, R. Wai, J. Tan, N. Ramesh, and A. Fluerasu for early \textit{in situ} work. X-ray scattering experiments and simulations were supported by the Office of Basic Energy Sciences (BES), US Department of Energy (DOE) (award no. DE-SC0019375). Use of the Stanford Synchrotron Radiation Lightsource, SLAC National Accelerator Laboratory, is supported by the DOE, Office of Science, Office of Basic Energy Sciences (contract no. DE-AC02-76SF00515). C.P.N.T. was supported by the NSF (Graduate Research Fellowship no. DGE1106400). J.K.U. was supported by an Arnold O. Beckman Postdoctoral Fellowship in Chemical Sciences from the Arnold and Mabel Beckman Foundation. A.D. was supported by a Philomathia Graduate Student Fellowship from the Kavli Nanoscience Institute at UC Berkeley. N.S.G., D.V.T. and D.T.L. were supported by Alfred P. Sloan Research Fellowships. N.S.G. and D.V.T. were also supported by  David and Lucile Packard Foundation Fellowships for Science and Engineering and Camille and Henry Dreyfus Teacher-Scholar Awards. 


\begin{suppinfo}

Experimental details, modeling of SAXS patterns, effects of polydispersity on the phase diagram, SL (111) FWHM power law exponents, simulation details, defect annealing in simulations. 

\end{suppinfo}

\section*{Associated content. }
A preprint version of this manuscript can be found in the arXiv preprint server.\cite{tanner_situ_2023}

\bibliography{main.bib}

\providecommand{\latin}[1]{#1}
\makeatletter
\providecommand{\doi}
  {\begingroup\let\do\@makeother\dospecials
  \catcode`\{=1 \catcode`\}=2 \doi@aux}
\providecommand{\doi@aux}[1]{\endgroup\texttt{#1}}
\makeatother
\providecommand*\mcitethebibliography{\thebibliography}
\csname @ifundefined\endcsname{endmcitethebibliography}  {\let\endmcitethebibliography\endthebibliography}{}
\begin{mcitethebibliography}{71}
\providecommand*\natexlab[1]{#1}
\providecommand*\mciteSetBstSublistMode[1]{}
\providecommand*\mciteSetBstMaxWidthForm[2]{}
\providecommand*\mciteBstWouldAddEndPuncttrue
  {\def\EndOfBibitem{\unskip.}}
\providecommand*\mciteBstWouldAddEndPunctfalse
  {\let\EndOfBibitem\relax}
\providecommand*\mciteSetBstMidEndSepPunct[3]{}
\providecommand*\mciteSetBstSublistLabelBeginEnd[3]{}
\providecommand*\EndOfBibitem{}
\mciteSetBstSublistMode{f}
\mciteSetBstMaxWidthForm{subitem}{(\alph{mcitesubitemcount})}
\mciteSetBstSublistLabelBeginEnd
  {\mcitemaxwidthsubitemform\space}
  {\relax}
  {\relax}

\bibitem[Boles \latin{et~al.}(2016)Boles, Engel, and Talapin]{boles_self-assembly_2016}
Boles,~M.~A.; Engel,~M.; Talapin,~D.~V. Self-{Assembly} of {Colloidal} {Nanocrystals}: {From} {Intricate} {Structures} to {Functional} {Materials}. \emph{Chemical Reviews} \textbf{2016}, \emph{116}, 11220--11289\relax
\mciteBstWouldAddEndPuncttrue
\mciteSetBstMidEndSepPunct{\mcitedefaultmidpunct}
{\mcitedefaultendpunct}{\mcitedefaultseppunct}\relax
\EndOfBibitem
\bibitem[Murray \latin{et~al.}(1995)Murray, Kagan, and Bawendi]{murray_self-organization_1995}
Murray,~C.~B.; Kagan,~C.~R.; Bawendi,~M.~G. Self-{Organization} of {CdSe} {Nanocrystallites} into {Three}-{Dimensional} {Quantum} {Dot} {Superlattices}. \emph{Science} \textbf{1995}, \emph{270}, 1335--1338\relax
\mciteBstWouldAddEndPuncttrue
\mciteSetBstMidEndSepPunct{\mcitedefaultmidpunct}
{\mcitedefaultendpunct}{\mcitedefaultseppunct}\relax
\EndOfBibitem
\bibitem[Shevchenko \latin{et~al.}(2006)Shevchenko, Talapin, Kotov, O'Brien, and Murray]{shevchenko_structural_2006}
Shevchenko,~E.~V.; Talapin,~D.~V.; Kotov,~N.~A.; O'Brien,~S.; Murray,~C.~B. Structural {Diversity} in {Binary} {Nanoparticle} {Superlattices}. \emph{Nature} \textbf{2006}, \emph{439}, 55--59\relax
\mciteBstWouldAddEndPuncttrue
\mciteSetBstMidEndSepPunct{\mcitedefaultmidpunct}
{\mcitedefaultendpunct}{\mcitedefaultseppunct}\relax
\EndOfBibitem
\bibitem[Smith \latin{et~al.}(2009)Smith, Goodfellow, Smilgies, and Korgel]{smith_self-assembled_2009}
Smith,~D.~K.; Goodfellow,~B.; Smilgies,~D.-M.; Korgel,~B.~A. Self-{Assembled} {Simple} {Hexagonal} {AB2} {Binary} {Nanocrystal} {Superlattices}: {SEM}, {GISAXS}, and {Defects}. \emph{Journal of the American Chemical Society} \textbf{2009}, \emph{131}, 3281--3290\relax
\mciteBstWouldAddEndPuncttrue
\mciteSetBstMidEndSepPunct{\mcitedefaultmidpunct}
{\mcitedefaultendpunct}{\mcitedefaultseppunct}\relax
\EndOfBibitem
\bibitem[Bian \latin{et~al.}(2011)Bian, Choi, Kaushik, Clancy, Smilgies, and Hanrath]{bian_shape-anisotropy_2011}
Bian,~K.; Choi,~J.~J.; Kaushik,~A.; Clancy,~P.; Smilgies,~D.-M.; Hanrath,~T. Shape-{Anisotropy} {Driven} {Symmetry} {Transformations} in {Nanocrystal} {Superlattice} {Polymorphs}. \emph{ACS Nano} \textbf{2011}, \emph{5}, 2815--2823\relax
\mciteBstWouldAddEndPuncttrue
\mciteSetBstMidEndSepPunct{\mcitedefaultmidpunct}
{\mcitedefaultendpunct}{\mcitedefaultseppunct}\relax
\EndOfBibitem
\bibitem[Wang \latin{et~al.}(2012)Wang, Wang, Breed, Manoharan, Feng, Hollingsworth, Weck, and Pine]{wang_colloids_2012}
Wang,~Y.; Wang,~Y.; Breed,~D.~R.; Manoharan,~V.~N.; Feng,~L.; Hollingsworth,~A.~D.; Weck,~M.; Pine,~D.~J. Colloids with {Valence} and {Specific} {Directional} {Bonding}. \emph{Nature} \textbf{2012}, \emph{491}, 51--55\relax
\mciteBstWouldAddEndPuncttrue
\mciteSetBstMidEndSepPunct{\mcitedefaultmidpunct}
{\mcitedefaultendpunct}{\mcitedefaultseppunct}\relax
\EndOfBibitem
\bibitem[Fan \latin{et~al.}(2010)Fan, Wu, Bao, Bao, Bardhan, Halas, Manoharan, Nordlander, Shvets, and Capasso]{fan_self-assembled_2010}
Fan,~J.~A.; Wu,~C.; Bao,~K.; Bao,~J.; Bardhan,~R.; Halas,~N.~J.; Manoharan,~V.~N.; Nordlander,~P.; Shvets,~G.; Capasso,~F. Self-{Assembled} {Plasmonic} {Nanoparticle} {Clusters}. \emph{Science} \textbf{2010}, \emph{328}, 1135--1138\relax
\mciteBstWouldAddEndPuncttrue
\mciteSetBstMidEndSepPunct{\mcitedefaultmidpunct}
{\mcitedefaultendpunct}{\mcitedefaultseppunct}\relax
\EndOfBibitem
\bibitem[Wang \latin{et~al.}(2012)Wang, Zhuang, Lynch, Chen, Wang, Wang, LaMontagne, Wu, Wang, and Cao]{wang_self-assembled_2012}
Wang,~T.; Zhuang,~J.; Lynch,~J.; Chen,~O.; Wang,~Z.; Wang,~X.; LaMontagne,~D.; Wu,~H.; Wang,~Z.; Cao,~Y.~C. Self-{Assembled} {Colloidal} {Superparticles} from {Nanorods}. \emph{Science} \textbf{2012}, \emph{338}, 358--363\relax
\mciteBstWouldAddEndPuncttrue
\mciteSetBstMidEndSepPunct{\mcitedefaultmidpunct}
{\mcitedefaultendpunct}{\mcitedefaultseppunct}\relax
\EndOfBibitem
\bibitem[Murray \latin{et~al.}(2000)Murray, Kagan, and Bawendi]{murray_synthesis_2000}
Murray,~C.~B.; Kagan,~C.~R.; Bawendi,~M.~G. Synthesis and {Characterization} of {Monodisperse} {Nanocrystals} and {Close}-{Packed} {Nanocrystal} {Assemblies}. \emph{Annual Review of Materials Science} \textbf{2000}, \emph{30}, 545--610\relax
\mciteBstWouldAddEndPuncttrue
\mciteSetBstMidEndSepPunct{\mcitedefaultmidpunct}
{\mcitedefaultendpunct}{\mcitedefaultseppunct}\relax
\EndOfBibitem
\bibitem[Santos \latin{et~al.}(2021)Santos, Gabrys, Zornberg, Lee, and Macfarlane]{santos_macroscopic_2021}
Santos,~P.~J.; Gabrys,~P.~A.; Zornberg,~L.~Z.; Lee,~M.~S.; Macfarlane,~R.~J. Macroscopic {Materials} {Assembled} from {Nanoparticle} {Superlattices}. \emph{Nature} \textbf{2021}, \emph{591}, 586--591\relax
\mciteBstWouldAddEndPuncttrue
\mciteSetBstMidEndSepPunct{\mcitedefaultmidpunct}
{\mcitedefaultendpunct}{\mcitedefaultseppunct}\relax
\EndOfBibitem
\bibitem[van Blaaderen \latin{et~al.}(1997)van Blaaderen, Ruel, and Wiltzius]{van_blaaderen_template-directed_1997}
van Blaaderen,~A.; Ruel,~R.; Wiltzius,~P. Template-{Directed} {Colloidal} {Crystallization}. \emph{Nature} \textbf{1997}, \emph{385}, 321--324\relax
\mciteBstWouldAddEndPuncttrue
\mciteSetBstMidEndSepPunct{\mcitedefaultmidpunct}
{\mcitedefaultendpunct}{\mcitedefaultseppunct}\relax
\EndOfBibitem
\bibitem[Gur \latin{et~al.}(2005)Gur, Fromer, Geier, and Alivisatos]{gur_air-stable_2005}
Gur,~I.; Fromer,~N.~A.; Geier,~M.~L.; Alivisatos,~A.~P. Air-{Stable} {All}-{Inorganic} {Nanocrystal} {Solar} {Cells} {Processed} from {Solution}. \emph{Science} \textbf{2005}, \emph{310}, 462--465\relax
\mciteBstWouldAddEndPuncttrue
\mciteSetBstMidEndSepPunct{\mcitedefaultmidpunct}
{\mcitedefaultendpunct}{\mcitedefaultseppunct}\relax
\EndOfBibitem
\bibitem[Wang \latin{et~al.}(2018)Wang, Wang, Huang, and Xu]{wang_materials_2018}
Wang,~C.; Wang,~C.; Huang,~Z.; Xu,~S. Materials and {Structures} toward {Soft} {Electronics}. \emph{Advanced Materials} \textbf{2018}, \emph{30}, 1801368\relax
\mciteBstWouldAddEndPuncttrue
\mciteSetBstMidEndSepPunct{\mcitedefaultmidpunct}
{\mcitedefaultendpunct}{\mcitedefaultseppunct}\relax
\EndOfBibitem
\bibitem[Arandiyan \latin{et~al.}(2015)Arandiyan, Dai, Ji, Sun, and Li]{arandiyan_pt_2015}
Arandiyan,~H.; Dai,~H.; Ji,~K.; Sun,~H.; Li,~J. Pt {Nanoparticles} {Embedded} in {Colloidal} {Crystal} {Template} {Derived} {3D} {Ordered} {Macroporous} {Ce}$_{\textrm{0.6}}${Zr}$_{\textrm{0.3}}${Y}$_{\textrm{0.1}}${O}$_{\textrm{2}}$: {Highly} {Efficient} {Catalysts} for {Methane} {Combustion}. \emph{ACS Catalysis} \textbf{2015}, \emph{5}, 1781--1793\relax
\mciteBstWouldAddEndPuncttrue
\mciteSetBstMidEndSepPunct{\mcitedefaultmidpunct}
{\mcitedefaultendpunct}{\mcitedefaultseppunct}\relax
\EndOfBibitem
\bibitem[Zhu and Hersam(2017)Zhu, and Hersam]{zhu_assembly_2017}
Zhu,~J.; Hersam,~M.~C. Assembly and {Electronic} {Applications} of {Colloidal} {Nanomaterials}. \emph{Advanced Materials} \textbf{2017}, \emph{29}, 1603895\relax
\mciteBstWouldAddEndPuncttrue
\mciteSetBstMidEndSepPunct{\mcitedefaultmidpunct}
{\mcitedefaultendpunct}{\mcitedefaultseppunct}\relax
\EndOfBibitem
\bibitem[Begley \latin{et~al.}(2019)Begley, Gianola, and Ray]{begley_bridging_2019}
Begley,~M.~R.; Gianola,~D.~S.; Ray,~T.~R. Bridging {Functional} {Nanocomposites} to {Robust} {Macroscale} {Devices}. \emph{Science} \textbf{2019}, \emph{364}, eaav4299\relax
\mciteBstWouldAddEndPuncttrue
\mciteSetBstMidEndSepPunct{\mcitedefaultmidpunct}
{\mcitedefaultendpunct}{\mcitedefaultseppunct}\relax
\EndOfBibitem
\bibitem[Israelachvili(2011)]{israelachvili_intermolecular_2011}
Israelachvili,~J. \emph{Intermolecular and {Surface} {Forces}}, 3rd ed.; Elsevier, Inc.: MA, 2011\relax
\mciteBstWouldAddEndPuncttrue
\mciteSetBstMidEndSepPunct{\mcitedefaultmidpunct}
{\mcitedefaultendpunct}{\mcitedefaultseppunct}\relax
\EndOfBibitem
\bibitem[Silvera~Batista \latin{et~al.}(2015)Silvera~Batista, Larson, and Kotov]{silvera_batista_nonadditivity_2015}
Silvera~Batista,~C.~A.; Larson,~R.~G.; Kotov,~N.~A. Nonadditivity of {Nanoparticle} {Interactions}. \emph{Science} \textbf{2015}, \emph{350}, 1242477\relax
\mciteBstWouldAddEndPuncttrue
\mciteSetBstMidEndSepPunct{\mcitedefaultmidpunct}
{\mcitedefaultendpunct}{\mcitedefaultseppunct}\relax
\EndOfBibitem
\bibitem[Pusey and van Megen(1986)Pusey, and van Megen]{pusey_phase_1986}
Pusey,~P.~N.; van Megen,~W. Phase {Behaviour} of {Concentrated} {Suspensions} of {Nearly} {Hard} {Colloidal} {Spheres}. \emph{Nature} \textbf{1986}, \emph{320}, 340--342, Number: 6060 Publisher: Nature Publishing Group\relax
\mciteBstWouldAddEndPuncttrue
\mciteSetBstMidEndSepPunct{\mcitedefaultmidpunct}
{\mcitedefaultendpunct}{\mcitedefaultseppunct}\relax
\EndOfBibitem
\bibitem[Gasser \latin{et~al.}(2001)Gasser, Weeks, Schofield, Pusey, and Weitz]{gasser_real-space_2001}
Gasser,~U.; Weeks,~E.~R.; Schofield,~A.; Pusey,~P.~N.; Weitz,~D.~A. Real-{Space} {Imaging} of {Nucleation} and {Growth} in {Colloidal} {Crystallization}. \emph{Science} \textbf{2001}, \emph{292}, 258--262, Publisher: American Association for the Advancement of Science\relax
\mciteBstWouldAddEndPuncttrue
\mciteSetBstMidEndSepPunct{\mcitedefaultmidpunct}
{\mcitedefaultendpunct}{\mcitedefaultseppunct}\relax
\EndOfBibitem
\bibitem[Calderon \latin{et~al.}(1993)Calderon, Bibette, and Biais]{calderon_experimental_1993}
Calderon,~F.~L.; Bibette,~J.; Biais,~J. Experimental {Phase} {Diagrams} of {Polymer} and {Colloid} {Mixtures}. \emph{Europhysics Letters} \textbf{1993}, \emph{23}, 653\relax
\mciteBstWouldAddEndPuncttrue
\mciteSetBstMidEndSepPunct{\mcitedefaultmidpunct}
{\mcitedefaultendpunct}{\mcitedefaultseppunct}\relax
\EndOfBibitem
\bibitem[Monovoukas and Gast(1989)Monovoukas, and Gast]{monovoukas_experimental_1989}
Monovoukas,~Y.; Gast,~A.~P. The {Experimental} {Phase} {Diagram} of {Charged} {Colloidal} {Suspensions}. \emph{Journal of Colloid and Interface Science} \textbf{1989}, \emph{128}, 533--548\relax
\mciteBstWouldAddEndPuncttrue
\mciteSetBstMidEndSepPunct{\mcitedefaultmidpunct}
{\mcitedefaultendpunct}{\mcitedefaultseppunct}\relax
\EndOfBibitem
\bibitem[Dinsmore \latin{et~al.}(1995)Dinsmore, Yodh, and Pine]{dinsmore_phase_1995}
Dinsmore,~A.~D.; Yodh,~A.~G.; Pine,~D.~J. Phase {Diagrams} of {Nearly}-{Hard}-{Sphere} {Binary} {Colloids}. \emph{Physical Review E} \textbf{1995}, \emph{52}, 4045--4057, Publisher: American Physical Society\relax
\mciteBstWouldAddEndPuncttrue
\mciteSetBstMidEndSepPunct{\mcitedefaultmidpunct}
{\mcitedefaultendpunct}{\mcitedefaultseppunct}\relax
\EndOfBibitem
\bibitem[Verhaegh \latin{et~al.}(1996)Verhaegh, van Duijneveldt, Dhont, and Lekkerkerker]{verhaegh_fluid-fluid_1996}
Verhaegh,~N. A.~M.; van Duijneveldt,~J.~S.; Dhont,~J. K.~G.; Lekkerkerker,~H. N.~W. Fluid-{Fluid} {Phase} {Separation} in {Colloid}-{Polymer} {Mixtures} {Studied} with {Small} {Angle} {Light} {Scattering} and {Light} {Microscopy}. \emph{Physica A: Statistical Mechanics and its Applications} \textbf{1996}, \emph{230}, 409--436\relax
\mciteBstWouldAddEndPuncttrue
\mciteSetBstMidEndSepPunct{\mcitedefaultmidpunct}
{\mcitedefaultendpunct}{\mcitedefaultseppunct}\relax
\EndOfBibitem
\bibitem[Weidman \latin{et~al.}(2016)Weidman, Smilgies, and Tisdale]{weidman_kinetics_2016}
Weidman,~M.~C.; Smilgies,~D.-M.; Tisdale,~W.~A. Kinetics of the {Self}-{Assembly} of {Nanocrystal} {Superlattices} {Measured} by {Real}-{Time} \textit{{In} {Situ}} {X}-ray {Scattering}. \emph{Nature Materials} \textbf{2016}, \emph{15}, 775--781\relax
\mciteBstWouldAddEndPuncttrue
\mciteSetBstMidEndSepPunct{\mcitedefaultmidpunct}
{\mcitedefaultendpunct}{\mcitedefaultseppunct}\relax
\EndOfBibitem
\bibitem[Graceffa \latin{et~al.}(2013)Graceffa, Nobrega, Barrea, Kathuria, Chakravarthy, Bilsel, and Irving]{graceffa_sub-millisecond_2013}
Graceffa,~R.; Nobrega,~R.~P.; Barrea,~R.~A.; Kathuria,~S.~V.; Chakravarthy,~S.; Bilsel,~O.; Irving,~T.~C. Sub-{Millisecond} {Time}-{Resolved} {SAXS} {Using} a {Continuous}-{Flow} {Mixer} and {X}-{Ray} {Microbeam}. \emph{Journal of Synchrotron Radiation} \textbf{2013}, \emph{20}, 820--825\relax
\mciteBstWouldAddEndPuncttrue
\mciteSetBstMidEndSepPunct{\mcitedefaultmidpunct}
{\mcitedefaultendpunct}{\mcitedefaultseppunct}\relax
\EndOfBibitem
\bibitem[Karim \latin{et~al.}(2015)Karim, Al~Hasan, Ivanov, Siefert, Kelly, Hallfors, Benavidez, Kovarik, Jenkins, Winans, and Datye]{karim_synthesis_2015}
Karim,~A.~M.; Al~Hasan,~N.; Ivanov,~S.; Siefert,~S.; Kelly,~R.~T.; Hallfors,~N.~G.; Benavidez,~A.; Kovarik,~L.; Jenkins,~A.; Winans,~R.~E.; Datye,~A.~K. Synthesis of 1 nm {Pd} {Nanoparticles} in a {Microfluidic} {Reactor}: {Insights} from \textit{{In} {Situ}} {X}-ray {Absorption} {Fine} {Structure} {Spectroscopy} and {Small}-{Angle} {X}-ray {Scattering}. \emph{The Journal of Physical Chemistry C} \textbf{2015}, \emph{119}, 13257--13267\relax
\mciteBstWouldAddEndPuncttrue
\mciteSetBstMidEndSepPunct{\mcitedefaultmidpunct}
{\mcitedefaultendpunct}{\mcitedefaultseppunct}\relax
\EndOfBibitem
\bibitem[Wu \latin{et~al.}(2018)Wu, Lian, Willis, Goodman, McKay, Qin, Tassone, and Cargnello]{wu_tuning_2018}
Wu,~L.; Lian,~H.; Willis,~J.~J.; Goodman,~E.~D.; McKay,~I.~S.; Qin,~J.; Tassone,~C.~J.; Cargnello,~M. Tuning {Precursor} {Reactivity} toward {Nanometer}-{Size} {Control} in {Palladium} {Nanoparticles} {Studied} by \textit{{In} {Situ}} {Small} {Angle} {X}-ray {Scattering}. \emph{Chemistry of Materials} \textbf{2018}, \emph{30}, 1127--1135\relax
\mciteBstWouldAddEndPuncttrue
\mciteSetBstMidEndSepPunct{\mcitedefaultmidpunct}
{\mcitedefaultendpunct}{\mcitedefaultseppunct}\relax
\EndOfBibitem
\bibitem[Korgel and Fitzmaurice(1999)Korgel, and Fitzmaurice]{korgel_small-angle_1999}
Korgel,~B.~A.; Fitzmaurice,~D. Small-{Angle} {X}-{Ray}-{Scattering} {Study} of {Silver}-{Nanocrystal} {Disorder}-{Order} {Phase} {Transitions}. \emph{Physical Review B} \textbf{1999}, \emph{59}, 14191--14201\relax
\mciteBstWouldAddEndPuncttrue
\mciteSetBstMidEndSepPunct{\mcitedefaultmidpunct}
{\mcitedefaultendpunct}{\mcitedefaultseppunct}\relax
\EndOfBibitem
\bibitem[Lu \latin{et~al.}(2012)Lu, Akey, Dahlman, Zhang, and Herman]{lu_resolving_2012}
Lu,~C.; Akey,~A.~J.; Dahlman,~C.~J.; Zhang,~D.; Herman,~I.~P. Resolving the {Growth} of {3D} {Colloidal} {Nanoparticle} {Superlattices} by {Real}-{Time} {Small}-{Angle} {X}-ray {Scattering}. \emph{Journal of the American Chemical Society} \textbf{2012}, \emph{134}, 18732--18738\relax
\mciteBstWouldAddEndPuncttrue
\mciteSetBstMidEndSepPunct{\mcitedefaultmidpunct}
{\mcitedefaultendpunct}{\mcitedefaultseppunct}\relax
\EndOfBibitem
\bibitem[Lokteva \latin{et~al.}(2021)Lokteva, Dartsch, Dallari, Westermeier, Walther, Grübel, and Lehmkühler]{lokteva_real-time_2021}
Lokteva,~I.; Dartsch,~M.; Dallari,~F.; Westermeier,~F.; Walther,~M.; Grübel,~G.; Lehmkühler,~F. Real-{Time} {X}-ray {Scattering} {Discovers} {Rich} {Phase} {Behavior} in {PbS} {Nanocrystal} {Superlattices} during \textit{{In} {Situ}} {Assembly}. \emph{Chemistry of Materials} \textbf{2021}, \emph{33}, 6553--6563\relax
\mciteBstWouldAddEndPuncttrue
\mciteSetBstMidEndSepPunct{\mcitedefaultmidpunct}
{\mcitedefaultendpunct}{\mcitedefaultseppunct}\relax
\EndOfBibitem
\bibitem[Josten \latin{et~al.}(2017)Josten, Wetterskog, Glavic, Boesecke, Feoktystov, Brauweiler-Reuters, Rücker, Salazar-Alvarez, Brückel, and Bergström]{josten_superlattice_2017}
Josten,~E.; Wetterskog,~E.; Glavic,~A.; Boesecke,~P.; Feoktystov,~A.; Brauweiler-Reuters,~E.; Rücker,~U.; Salazar-Alvarez,~G.; Brückel,~T.; Bergström,~L. Superlattice {Growth} and {Rearrangement} during {Evaporation}-{Induced} {Nanoparticle} {Self}-{Assembly}. \emph{Scientific Reports} \textbf{2017}, \emph{7}, 2802\relax
\mciteBstWouldAddEndPuncttrue
\mciteSetBstMidEndSepPunct{\mcitedefaultmidpunct}
{\mcitedefaultendpunct}{\mcitedefaultseppunct}\relax
\EndOfBibitem
\bibitem[Smilgies and Hanrath(2017)Smilgies, and Hanrath]{smilgies_superlattice_2017}
Smilgies,~D.-M.; Hanrath,~T. Superlattice {Self}-{Assembly}: {Watching} {Nanocrystals} in {Action}(a). \emph{Europhysics Letters} \textbf{2017}, \emph{119}, 28003\relax
\mciteBstWouldAddEndPuncttrue
\mciteSetBstMidEndSepPunct{\mcitedefaultmidpunct}
{\mcitedefaultendpunct}{\mcitedefaultseppunct}\relax
\EndOfBibitem
\bibitem[Geuchies \latin{et~al.}(2016)Geuchies, van Overbeek, Evers, Goris, de~Backer, Gantapara, Rabouw, Hilhorst, Peters, Konovalov, Petukhov, Dijkstra, Siebbeles, van Aert, Bals, and Vanmaekelbergh]{geuchies_situ_2016}
Geuchies,~J.~J. \latin{et~al.}  \textit{{In} {Situ}} {Study} of the {Formation} {Mechanism} of {Two}-{Dimensional} {Superlattices} from {PbSe} {Nanocrystals}. \emph{Nature Materials} \textbf{2016}, \emph{15}, 1248--1254\relax
\mciteBstWouldAddEndPuncttrue
\mciteSetBstMidEndSepPunct{\mcitedefaultmidpunct}
{\mcitedefaultendpunct}{\mcitedefaultseppunct}\relax
\EndOfBibitem
\bibitem[Wu \latin{et~al.}(2017)Wu, Willis, McKay, Diroll, Qin, Cargnello, and Tassone]{wu_high-temperature_2017}
Wu,~L.; Willis,~J.~J.; McKay,~I.~S.; Diroll,~B.~T.; Qin,~J.; Cargnello,~M.; Tassone,~C.~J. High-{Temperature} {Crystallization} of {Nanocrystals} into {Three}-{Dimensional} {Superlattices}. \emph{Nature} \textbf{2017}, \emph{548}, 197--201\relax
\mciteBstWouldAddEndPuncttrue
\mciteSetBstMidEndSepPunct{\mcitedefaultmidpunct}
{\mcitedefaultendpunct}{\mcitedefaultseppunct}\relax
\EndOfBibitem
\bibitem[Abécassis \latin{et~al.}(2008)Abécassis, Testard, and Spalla]{abecassis_gold_2008}
Abécassis,~B.; Testard,~F.; Spalla,~O. Gold {Nanoparticle} {Superlattice} {Crystallization} {Probed} \textit{{In} {Situ}}. \emph{Physical Review Letters} \textbf{2008}, \emph{100}, 115504\relax
\mciteBstWouldAddEndPuncttrue
\mciteSetBstMidEndSepPunct{\mcitedefaultmidpunct}
{\mcitedefaultendpunct}{\mcitedefaultseppunct}\relax
\EndOfBibitem
\bibitem[Marino \latin{et~al.}(2023)Marino, Rosen, Yang, Tsai, and Murray]{marino_temperature-controlled_2023}
Marino,~E.; Rosen,~D.~J.; Yang,~S.; Tsai,~E.~H.; Murray,~C.~B. Temperature-{Controlled} {Reversible} {Formation} and {Phase} {Transformation} of {3D} {Nanocrystal} {Superlattices} {Through} \textit{{In} {Situ}} {Small}-{Angle} {X}-ray {Scattering}. \emph{Nano Letters} \textbf{2023}, \emph{23}, 4250--4257\relax
\mciteBstWouldAddEndPuncttrue
\mciteSetBstMidEndSepPunct{\mcitedefaultmidpunct}
{\mcitedefaultendpunct}{\mcitedefaultseppunct}\relax
\EndOfBibitem
\bibitem[Grote \latin{et~al.}(2021)Grote, Zito, Frank, Dippel, Reisbeck, Pitala, Kvashnina, Bauters, Detlefs, Ivashko, Pandit, Rebber, Harouna-Mayer, Nickel, and Koziej]{grote_x-ray_2021}
Grote,~L.; Zito,~C.~A.; Frank,~K.; Dippel,~A.-C.; Reisbeck,~P.; Pitala,~K.; Kvashnina,~K.~O.; Bauters,~S.; Detlefs,~B.; Ivashko,~O.; Pandit,~P.; Rebber,~M.; Harouna-Mayer,~S.~Y.; Nickel,~B.; Koziej,~D. X-ray {Studies} {Bridge} the {Molecular} and {Macro} {Length} {Scales} during the {Emergence} of {CoO} {Assemblies}. \emph{Nature Communications} \textbf{2021}, \emph{12}, 4429\relax
\mciteBstWouldAddEndPuncttrue
\mciteSetBstMidEndSepPunct{\mcitedefaultmidpunct}
{\mcitedefaultendpunct}{\mcitedefaultseppunct}\relax
\EndOfBibitem
\bibitem[Qiao \latin{et~al.}(2023)Qiao, Wang, Zhai, Yu, Fang, and Chen]{qiao_situ_2023}
Qiao,~Z.; Wang,~X.; Zhai,~Y.; Yu,~R.; Fang,~Z.; Chen,~G. \textit{{In} {Situ}} {Real}-{Time} {Observation} of {Formation} and {Self}-{Assembly} of {Perovskite} {Nanocrystals} at {High} {Temperature}. \emph{Nano Letters} \textbf{2023}, \emph{23}, 10788--10795\relax
\mciteBstWouldAddEndPuncttrue
\mciteSetBstMidEndSepPunct{\mcitedefaultmidpunct}
{\mcitedefaultendpunct}{\mcitedefaultseppunct}\relax
\EndOfBibitem
\bibitem[Kovalenko \latin{et~al.}(2009)Kovalenko, Scheele, and Talapin]{kovalenko_colloidal_2009}
Kovalenko,~M.~V.; Scheele,~M.; Talapin,~D.~V. Colloidal {Nanocrystals} with {Molecular} {Metal} {Chalcogenide} {Surface} {Ligands}. \emph{Science} \textbf{2009}, \emph{324}, 1417--1420\relax
\mciteBstWouldAddEndPuncttrue
\mciteSetBstMidEndSepPunct{\mcitedefaultmidpunct}
{\mcitedefaultendpunct}{\mcitedefaultseppunct}\relax
\EndOfBibitem
\bibitem[Coropceanu \latin{et~al.}(2022)Coropceanu, Janke, Portner, Haubold, Nguyen, Das, Tanner, Utterback, Teitelbaum, Hudson, Sarma, Hinkle, Tassone, Eychmüller, Limmer, Olvera de~la Cruz, Ginsberg, and Talapin]{coropceanu_self-assembly_2022}
Coropceanu,~I. \latin{et~al.}  Self-assembly of {Nanocrystals} into {Strongly} {Electronically} {Coupled} {All}-{Inorganic} {Supercrystals}. \emph{Science} \textbf{2022}, \emph{375}, 1422--1426\relax
\mciteBstWouldAddEndPuncttrue
\mciteSetBstMidEndSepPunct{\mcitedefaultmidpunct}
{\mcitedefaultendpunct}{\mcitedefaultseppunct}\relax
\EndOfBibitem
\bibitem[Warren(1990)]{warren_x-ray_1990}
Warren,~B. \emph{X-ray {Diffraction}}, 2nd ed.; Dover Publications, Inc.: NY, 1990\relax
\mciteBstWouldAddEndPuncttrue
\mciteSetBstMidEndSepPunct{\mcitedefaultmidpunct}
{\mcitedefaultendpunct}{\mcitedefaultseppunct}\relax
\EndOfBibitem
\bibitem[ten Wolde and Frenkel(1997)ten Wolde, and Frenkel]{ten_wolde_enhancement_1997}
ten Wolde,~P.~R.; Frenkel,~D. Enhancement of {Protein} {Crystal} {Nucleation} by {Critical} {Density} {Fluctuations}. \emph{Science} \textbf{1997}, \emph{277}, 1975--1978\relax
\mciteBstWouldAddEndPuncttrue
\mciteSetBstMidEndSepPunct{\mcitedefaultmidpunct}
{\mcitedefaultendpunct}{\mcitedefaultseppunct}\relax
\EndOfBibitem
\bibitem[Wedekind \latin{et~al.}(2015)Wedekind, Xu, Buldyrev, Stanley, Reguera, and Franzese]{wedekind_optimization_2015}
Wedekind,~J.; Xu,~L.; Buldyrev,~S.~V.; Stanley,~H.~E.; Reguera,~D.; Franzese,~G. Optimization of {Crystal} {Nucleation} {Close} to a {Metastable} {Fluid}-{Fluid} {Phase} {Transition}. \emph{Scientific Reports} \textbf{2015}, \emph{5}, 11260\relax
\mciteBstWouldAddEndPuncttrue
\mciteSetBstMidEndSepPunct{\mcitedefaultmidpunct}
{\mcitedefaultendpunct}{\mcitedefaultseppunct}\relax
\EndOfBibitem
\bibitem[Haxton \latin{et~al.}(2015)Haxton, Hedges, and Whitelam]{haxton_crystallization_2015}
Haxton,~T.~K.; Hedges,~L.~O.; Whitelam,~S. Crystallization and {Arrest} {Mechanisms} of {Model} {Colloids}. \emph{Soft Matter} \textbf{2015}, \emph{11}, 9307--9320\relax
\mciteBstWouldAddEndPuncttrue
\mciteSetBstMidEndSepPunct{\mcitedefaultmidpunct}
{\mcitedefaultendpunct}{\mcitedefaultseppunct}\relax
\EndOfBibitem
\bibitem[Karthika \latin{et~al.}(2016)Karthika, Radhakrishnan, and Kalaichelvi]{karthika_review_2016}
Karthika,~S.; Radhakrishnan,~T.~K.; Kalaichelvi,~P. A {Review} of {Classical} and {Nonclassical} {Nucleation} {Theories}. \emph{Crystal Growth \& Design} \textbf{2016}, \emph{16}, 6663--6681\relax
\mciteBstWouldAddEndPuncttrue
\mciteSetBstMidEndSepPunct{\mcitedefaultmidpunct}
{\mcitedefaultendpunct}{\mcitedefaultseppunct}\relax
\EndOfBibitem
\bibitem[Savage and Dinsmore(2009)Savage, and Dinsmore]{savage_experimental_2009}
Savage,~J.~R.; Dinsmore,~A.~D. Experimental {Evidence} for {Two}-{Step} {Nucleation} in {Colloidal} {Crystallization}. \emph{Physical Review Letters} \textbf{2009}, \emph{102}, 198302\relax
\mciteBstWouldAddEndPuncttrue
\mciteSetBstMidEndSepPunct{\mcitedefaultmidpunct}
{\mcitedefaultendpunct}{\mcitedefaultseppunct}\relax
\EndOfBibitem
\bibitem[Zhang \latin{et~al.}(2012)Zhang, Roth, Wolf, Roosen-Runge, A. Skoda, J. Jacobs, Stzucki, and Schreiber]{zhang_charge-controlled_2012}
Zhang,~F.; Roth,~R.; Wolf,~M.; Roosen-Runge,~F.; A. Skoda,~M.~W.; J. Jacobs,~R.~M.; Stzucki,~M.; Schreiber,~F. Charge-{Controlled} {Metastable} {Liquid}–{Liquid} {Phase} {Separation} in {Protein} {Solutions} {As} a {Universal} {Pathway} towards {Crystallization}. \emph{Soft Matter} \textbf{2012}, \emph{8}, 1313--1316\relax
\mciteBstWouldAddEndPuncttrue
\mciteSetBstMidEndSepPunct{\mcitedefaultmidpunct}
{\mcitedefaultendpunct}{\mcitedefaultseppunct}\relax
\EndOfBibitem
\bibitem[Pagan and Gunton(2005)Pagan, and Gunton]{pagan_phase_2005}
Pagan,~D.~L.; Gunton,~J.~D. Phase {Behavior} of {Short}-{Range} {Square}-{Well} {Model}. \emph{The Journal of Chemical Physics} \textbf{2005}, \emph{122}, 184515\relax
\mciteBstWouldAddEndPuncttrue
\mciteSetBstMidEndSepPunct{\mcitedefaultmidpunct}
{\mcitedefaultendpunct}{\mcitedefaultseppunct}\relax
\EndOfBibitem
\bibitem[Vekilov(2010)]{vekilov_two-step_2010}
Vekilov,~P.~G. The {Two}-{Step} {Mechanism} of {Nucleation} of {Crystals} in {Solution}. \emph{Nanoscale} \textbf{2010}, \emph{2}, 2346--2357\relax
\mciteBstWouldAddEndPuncttrue
\mciteSetBstMidEndSepPunct{\mcitedefaultmidpunct}
{\mcitedefaultendpunct}{\mcitedefaultseppunct}\relax
\EndOfBibitem
\bibitem[Lee \latin{et~al.}(2019)Lee, Teich, Engel, and Glotzer]{lee_entropic_2019}
Lee,~S.; Teich,~E.~G.; Engel,~M.; Glotzer,~S.~C. Entropic {Colloidal} {Crystallization} {Pathways} via {Fluid}–{Fluid} {Transitions} and {Multidimensional} {Prenucleation} {Motifs}. \emph{Proceedings of the National Academy of Sciences} \textbf{2019}, \emph{116}, 14843--14851\relax
\mciteBstWouldAddEndPuncttrue
\mciteSetBstMidEndSepPunct{\mcitedefaultmidpunct}
{\mcitedefaultendpunct}{\mcitedefaultseppunct}\relax
\EndOfBibitem
\bibitem[Noro and Frenkel(2000)Noro, and Frenkel]{noro_extended_2000}
Noro,~M.~G.; Frenkel,~D. Extended {Corresponding}-{States} {Behavior} for {Particles} with {Variable} {Range} {Attractions}. \emph{The Journal of Chemical Physics} \textbf{2000}, \emph{113}, 2941--2944\relax
\mciteBstWouldAddEndPuncttrue
\mciteSetBstMidEndSepPunct{\mcitedefaultmidpunct}
{\mcitedefaultendpunct}{\mcitedefaultseppunct}\relax
\EndOfBibitem
\bibitem[Steinhardt \latin{et~al.}(1983)Steinhardt, Nelson, and Ronchetti]{steinhardt_bond-orientational_1983}
Steinhardt,~P.~J.; Nelson,~D.~R.; Ronchetti,~M. Bond-{Orientational} {Order} in {Liquids} and {Glasses}. \emph{Physical Review B} \textbf{1983}, \emph{28}, 784--805\relax
\mciteBstWouldAddEndPuncttrue
\mciteSetBstMidEndSepPunct{\mcitedefaultmidpunct}
{\mcitedefaultendpunct}{\mcitedefaultseppunct}\relax
\EndOfBibitem
\bibitem[Klotsa and Jack(2011)Klotsa, and Jack]{klotsa_predicting_2011}
Klotsa,~D.; Jack,~R.~L. Predicting the {Self}-assembly of a {Model} {Colloidal} {Crystal}. \emph{Soft Matter} \textbf{2011}, \emph{7}, 6294--6303\relax
\mciteBstWouldAddEndPuncttrue
\mciteSetBstMidEndSepPunct{\mcitedefaultmidpunct}
{\mcitedefaultendpunct}{\mcitedefaultseppunct}\relax
\EndOfBibitem
\bibitem[Hurley \latin{et~al.}()Hurley, Tanner, Portner, Utterback, Coropceanu, Williams, Das, Fluerasu, Sun, Song, Hamerlynck, Miller, Bhattacharyya, Talapin, Ginsberg, and Teitelbaum]{hurley_situ_2024}
Hurley,~M.~J. \latin{et~al.}  \textit{{In} {Situ}} {Coherent} {X}-ray {Scattering} {Reveals} {Polycrystalline} {Structure} and {Discrete} {Annealing} {Events} in {Strongly}-{Coupled} {Nanocrystal} {Superlattices}. \textbf{2024}. arXiv:2401.06103. arXiv Preprint (Condensed Matter, Materials Science, Mesoscale and Nanoscale Physics, Soft Condensed Matter), http://arxiv.org/abs/2401.06103 (accessed 01/25/2024.)\relax
\mciteBstWouldAddEndPuncttrue
\mciteSetBstMidEndSepPunct{\mcitedefaultmidpunct}
{\mcitedefaultendpunct}{\mcitedefaultseppunct}\relax
\EndOfBibitem
\bibitem[Schöpe \latin{et~al.}(2007)Schöpe, Bryant, and van Megen]{schope_effect_2007}
Schöpe,~H.~J.; Bryant,~G.; van Megen,~W. Effect of {Polydispersity} on the {Crystallization} {Kinetics} of {Suspensions} of {Colloidal} {Hard} {Spheres} when {Approaching} the {Glass} {Transition}. \emph{The Journal of Chemical Physics} \textbf{2007}, \emph{127}, 084505\relax
\mciteBstWouldAddEndPuncttrue
\mciteSetBstMidEndSepPunct{\mcitedefaultmidpunct}
{\mcitedefaultendpunct}{\mcitedefaultseppunct}\relax
\EndOfBibitem
\bibitem[Auer and Frenkel(2001)Auer, and Frenkel]{auer_prediction_2001}
Auer,~S.; Frenkel,~D. Prediction of {Absolute} {Crystal}-{Nucleation} {Rate} in {Hard}-{Sphere} {Colloids}. \emph{Nature} \textbf{2001}, \emph{409}, 1020--1023\relax
\mciteBstWouldAddEndPuncttrue
\mciteSetBstMidEndSepPunct{\mcitedefaultmidpunct}
{\mcitedefaultendpunct}{\mcitedefaultseppunct}\relax
\EndOfBibitem
\bibitem[Galkin and Vekilov(2000)Galkin, and Vekilov]{galkin_control_2000}
Galkin,~O.; Vekilov,~P.~G. Control of {Protein} {Crystal} {Nucleation} {Around} the {Metastable} {Liquid}–{Liquid} {Phase} {Boundary}. \emph{Proceedings of the National Academy of Sciences} \textbf{2000}, \emph{97}, 6277--6281\relax
\mciteBstWouldAddEndPuncttrue
\mciteSetBstMidEndSepPunct{\mcitedefaultmidpunct}
{\mcitedefaultendpunct}{\mcitedefaultseppunct}\relax
\EndOfBibitem
\bibitem[Seoane \latin{et~al.}(2016)Seoane, Castellanos, Dikhtiarenko, Kapteijn, and Gascon]{seoane_multi-scale_2016}
Seoane,~B.; Castellanos,~S.; Dikhtiarenko,~A.; Kapteijn,~F.; Gascon,~J. Multi-scale {Crystal} {Engineering} of {Metal} {Organic} {Frameworks}. \emph{Coordination Chemistry Reviews} \textbf{2016}, \emph{307}, 147--187\relax
\mciteBstWouldAddEndPuncttrue
\mciteSetBstMidEndSepPunct{\mcitedefaultmidpunct}
{\mcitedefaultendpunct}{\mcitedefaultseppunct}\relax
\EndOfBibitem
\bibitem[Millange \latin{et~al.}(2010)Millange, Medina, Guillou, Férey, Golden, and Walton]{millange_time-resolved_2010}
Millange,~F.; Medina,~M.~I.; Guillou,~N.; Férey,~G.; Golden,~K.~M.; Walton,~R.~I. Time-{Resolved} \textit{{In} {Situ}} {Diffraction} {Study} of the {Solvothermal} {Crystallization} of {Some} {Prototypical} {Metal}–{Organic} {Frameworks}. \emph{Angewandte Chemie} \textbf{2010}, \emph{122}, 775--778\relax
\mciteBstWouldAddEndPuncttrue
\mciteSetBstMidEndSepPunct{\mcitedefaultmidpunct}
{\mcitedefaultendpunct}{\mcitedefaultseppunct}\relax
\EndOfBibitem
\bibitem[Cheetham \latin{et~al.}(2018)Cheetham, Kieslich, and Yeung]{cheetham_thermodynamic_2018}
Cheetham,~A.~K.; Kieslich,~G.; Yeung,~H. H.-M. Thermodynamic and {Kinetic} {Effects} in the {Crystallization} of {Metal}–{Organic} {Frameworks}. \emph{Accounts of Chemical Research} \textbf{2018}, \emph{51}, 659--667\relax
\mciteBstWouldAddEndPuncttrue
\mciteSetBstMidEndSepPunct{\mcitedefaultmidpunct}
{\mcitedefaultendpunct}{\mcitedefaultseppunct}\relax
\EndOfBibitem
\bibitem[Feng \latin{et~al.}(2012)Feng, Ding, and Jiang]{feng_covalent_2012}
Feng,~X.; Ding,~X.; Jiang,~D. Covalent {Organic} {Frameworks}. \emph{Chemical Society Reviews} \textbf{2012}, \emph{41}, 6010--6022\relax
\mciteBstWouldAddEndPuncttrue
\mciteSetBstMidEndSepPunct{\mcitedefaultmidpunct}
{\mcitedefaultendpunct}{\mcitedefaultseppunct}\relax
\EndOfBibitem
\bibitem[Li \latin{et~al.}(2017)Li, Chavez, Li, Li, Dichtel, and Bredas]{li_nucleation_2017}
Li,~H.; Chavez,~A.~D.; Li,~H.; Li,~H.; Dichtel,~W.~R.; Bredas,~J.-L. Nucleation and {Growth} of {Covalent} {Organic} {Frameworks} from {Solution}: {The} {Example} of {COF}-5. \emph{Journal of the American Chemical Society} \textbf{2017}, \emph{139}, 16310--16318\relax
\mciteBstWouldAddEndPuncttrue
\mciteSetBstMidEndSepPunct{\mcitedefaultmidpunct}
{\mcitedefaultendpunct}{\mcitedefaultseppunct}\relax
\EndOfBibitem
\bibitem[Smith and Dichtel(2014)Smith, and Dichtel]{smith_mechanistic_2014}
Smith,~B.~J.; Dichtel,~W.~R. Mechanistic {Studies} of {Two}-{Dimensional} {Covalent} {Organic} {Frameworks} {Rapidly} {Polymerized} from {Initially} {Homogenous} {Conditions}. \emph{Journal of the American Chemical Society} \textbf{2014}, \emph{136}, 8783--8789\relax
\mciteBstWouldAddEndPuncttrue
\mciteSetBstMidEndSepPunct{\mcitedefaultmidpunct}
{\mcitedefaultendpunct}{\mcitedefaultseppunct}\relax
\EndOfBibitem
\bibitem[Song \latin{et~al.}(2013)Song, Wang, and Ding]{song_smart_2013}
Song,~C.; Wang,~Z.-G.; Ding,~B. Smart {Nanomachines} {Based} on {DNA} {Self}-{Assembly}. \emph{Small} \textbf{2013}, \emph{9}, 2382--2392\relax
\mciteBstWouldAddEndPuncttrue
\mciteSetBstMidEndSepPunct{\mcitedefaultmidpunct}
{\mcitedefaultendpunct}{\mcitedefaultseppunct}\relax
\EndOfBibitem
\bibitem[Shin \latin{et~al.}(2020)Shin, Shukla, Chung, Beiss, Chan, Ortega-Rivera, Wirth, Chen, Sack, Pokorski, and Steinmetz]{shin_covid-19_2020}
Shin,~M.~D.; Shukla,~S.; Chung,~Y.~H.; Beiss,~V.; Chan,~S.~K.; Ortega-Rivera,~O.~A.; Wirth,~D.~M.; Chen,~A.; Sack,~M.; Pokorski,~J.~K.; Steinmetz,~N.~F. {COVID}-19 {Vaccine} {Development} and a {Potential} {Nanomaterial} {Path} {Forward}. \emph{Nature Nanotechnology} \textbf{2020}, \emph{15}, 646--655\relax
\mciteBstWouldAddEndPuncttrue
\mciteSetBstMidEndSepPunct{\mcitedefaultmidpunct}
{\mcitedefaultendpunct}{\mcitedefaultseppunct}\relax
\EndOfBibitem
\bibitem[Biggs and Mulvaney(1994)Biggs, and Mulvaney]{biggs_measurement_1994}
Biggs,~S.; Mulvaney,~P. Measurement of the {Forces} between {Gold} {Surfaces} in {Water} by {Atomic} {Force} {Microscopy}. \emph{The Journal of Chemical Physics} \textbf{1994}, \emph{100}, 8501--8505\relax
\mciteBstWouldAddEndPuncttrue
\mciteSetBstMidEndSepPunct{\mcitedefaultmidpunct}
{\mcitedefaultendpunct}{\mcitedefaultseppunct}\relax
\EndOfBibitem
\bibitem[Thompson \latin{et~al.}(2022)Thompson, Aktulga, Berger, Bolintineanu, Brown, Crozier, in~'t Veld, Kohlmeyer, Moore, Nguyen, Shan, Stevens, Tranchida, Trott, and Plimpton]{thompson_lammps_2022}
Thompson,~A.~P.; Aktulga,~H.~M.; Berger,~R.; Bolintineanu,~D.~S.; Brown,~W.~M.; Crozier,~P.~S.; in~'t Veld,~P.~J.; Kohlmeyer,~A.; Moore,~S.~G.; Nguyen,~T.~D.; Shan,~R.; Stevens,~M.~J.; Tranchida,~J.; Trott,~C.; Plimpton,~S.~J. {LAMMPS} - a {Flexible} {Simulation} {Tool} for {Particle}-{Based} {Materials} {Modeling} at the {Atomic}, {Meso}, and {Continuum} {Scales}. \emph{Computer Physics Communications} \textbf{2022}, \emph{271}, 108171\relax
\mciteBstWouldAddEndPuncttrue
\mciteSetBstMidEndSepPunct{\mcitedefaultmidpunct}
{\mcitedefaultendpunct}{\mcitedefaultseppunct}\relax
\EndOfBibitem
\bibitem[Weeks \latin{et~al.}(1971)Weeks, Chandler, and Andersen]{weeks_role_1971}
Weeks,~J.~D.; Chandler,~D.; Andersen,~H.~C. Role of {Repulsive} {Forces} in {Determining} the {Equilibrium} {Structure} of {Simple} {Liquids}. \emph{The Journal of Chemical Physics} \textbf{1971}, \emph{54}, 5237--5247\relax
\mciteBstWouldAddEndPuncttrue
\mciteSetBstMidEndSepPunct{\mcitedefaultmidpunct}
{\mcitedefaultendpunct}{\mcitedefaultseppunct}\relax
\EndOfBibitem
\bibitem[Tanner \latin{et~al.}()Tanner, Utterback, Portner, Coropceanu, Das, Tassone, Teitelbaum, Limmer, Talapin, and Ginsberg]{tanner_situ_2023}
Tanner,~C. P.~N.; Utterback,~J.~K.; Portner,~J.; Coropceanu,~I.; Das,~A.; Tassone,~C.~J.; Teitelbaum,~S.~W.; Limmer,~D.~T.; Talapin,~D.~V.; Ginsberg,~N.~S. \textit{In} \textit{situ} {X}-ray scattering reveals coarsening rates of superlattices self-assembled from electrostatically stabilized metal nanocrystals depend non-monotonically on driving force. \textbf{2023}. arXiv:2312.06852. arXiv Preprint (Condensed Matter, Soft Condensed Matter), http://arxiv.org/abs/2312.06852 (accessed 01/25/2024.)\relax
\mciteBstWouldAddEndPuncttrue
\mciteSetBstMidEndSepPunct{\mcitedefaultmidpunct}
{\mcitedefaultendpunct}{\mcitedefaultseppunct}\relax
\EndOfBibitem
\end{mcitethebibliography}

\begin{tocentry}
\begin{center}
\includegraphics[width=7cm]{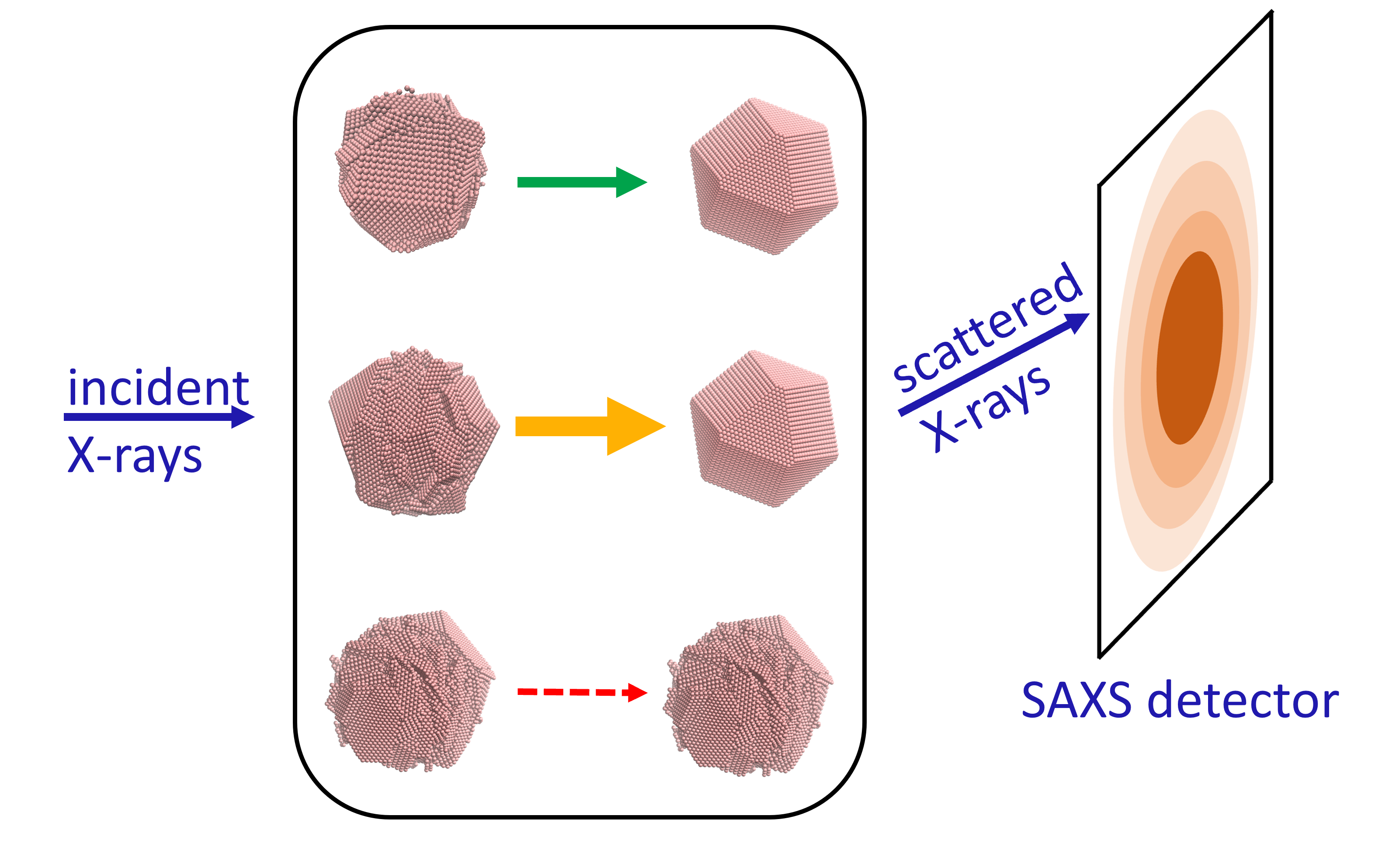}
\end{center}
\end{tocentry}

\end{document}


\pagebreak
\singlespacing

\textbf{Nanocrystal surface chemistry.} The nanocrystal solvation shell is currently under investigation because its characteristics are more complex than a typical double layer. \cite{zhang_stable_2017,kamysbayev_nanocrystals_2019,silvera_batista_nonadditivity_2015} The $\zeta$-potential for the type of NCs used in this study was previously measured\cite{kovalenko_colloidal_2009} to be $\sim$ -0.4 V. We do not expect these Au NCs to be faceted due to their small size. With the addition of excess (N$_2$H$_5$)Sn$_2$S$_6$ salt, we do not expect the surface charge to change but we do expect the concentration of ions near the NC surface to increase based on simulation work.\cite{guerrero_garcia_polarization_2014} Self-assembly of NCs into ordered SLs requires the addition of a salt containing multivalent ions.\cite{coropceanu_self-assembly_2022}

\textbf{Limitations of DLVO theory.} DLVO theory treats ions in solution as point particles. This assumption breaks down for NCs whose diameters are comparable in size to the ions themselves\cite{silvera_batista_nonadditivity_2015}, which is the case for the Au NCs (4.5 nm) and Sn$_2$S$_6^{4-}$ ions ($\sim$0.7 nm) in this work. The sizes of the ions are also comparable to and even larger than the calculated Debye lengths ($\sim$0.14-0.27 nm). In addition, the Hamaker constant used in the DLVO calculations in this work is taken from measurements between (flat) gold surfaces in water\cite{biggs_measurement_1994} since accurate measurements for NCs are lacking. 

\textbf{Temperature stability of the apparatus.} The \textit{in situ} apparatus is not thermally stabilized, however, we do not expect the temperature of the apparatus to change by more than a few degrees C (or $\sim$0.01 $k_\mathrm{B}T$, which is negligible on the scales of the well depths) at most due to fluctuations in room temperature and absorption of X-rays based on the X-ray flux, beam size, and the specific heat of the sample. 



\textbf{Modeling of SAXS patterns.} We model the background-subtracted scattered intensity as $I(q) = I_\mathrm{colloid}(q) + I_\mathrm{SL}(q)$, where $I_\mathrm{colloid}(q)$ is the scattered intensity from dilute, polydisperse, hard spheres and $I_\mathrm{SL}(q)$ is the scattered intensity from finite-sized \textit{fcc} SLs. Although the NCs have surface ligands, the hard sphere approximation is a good one because the scattering from the Au NCs is much stronger than the scattering from the ligands, which are composed of lighter elements. In addition, any faceting of these NCs is not substantial enough to change the form of the measured SAXS patterns. Typically, $I_\mathrm{colloid}(q) = P(q)$, where $P(q)$ is the form factor for hard spheres of radius $r$ and takes the form $P(q,r,\sigma,c_1) =c_1\int_0^\infty |F(q,r)|^2V(r)^2P(r,\sigma)dr$, where $F(q,r) = 3\frac{\sin(qr) - qr\cos(qr)}{(qr)^3}$, $V(r)$ is the volume of the hard sphere, $P(r,\sigma)$ is a Gaussian probability distribution centered at $r$ with standard deviation $\sigma$, and $c_1$ is a number directly proportional to the number of hard spheres. To compute $P(q)$ we use xrsdkit (https://github.com/scattering-central/xrsdkit). For the SL term, we use  $I_\mathrm{SL}(q) = P(q,r,\sigma,c_2)S_\mathrm{SL}(q)$, where $S_\mathrm{SL}(q)$ is the structure factor for a finite-sized \textit{fcc} SL. The structure factor describes the modulation in $q$ of the scattered intensity due to the spatial arrangement of the NCs. We model $S_\mathrm{SL}(q)$ as the sum of a set of Lorentzian line shapes each centered on a respective Bragg peak of an \textit{fcc} lattice (indexed by their Miller indices) and an additional $q^{-4}$ term. The Lorentzian lineshape is defined as $S_\mathrm{L}(q,q_0,\Gamma) = \frac{1}{\pi}\frac{\frac{1}{2}\Gamma}{(q-q_0)^2+(\frac{1}{2}\Gamma)^2}$, where $q_0$ is the center of the curve and $\Gamma$ is the full width at half maximum (FWHM). There is an additional side peak that appears at slightly lower $q$ than the \textit{fcc} (111) peak that we also model with a Lorentzian lineshape and index with the letters $\mathrm{lq}$. The final model for the SL structure factor is $S_\mathrm{SL}(q) =c_\mathrm{lq}S_\mathrm{L}(q,q_\mathrm{lq},\Gamma_\mathrm{lq})+c_{111}S_\mathrm{L}(q,q_{111},\Gamma_{111})+c_{200}S_\mathrm{L}(q,q_{200},\Gamma_{200}) 
+c_{220}S_\mathrm{L}(q,q_{220},\Gamma_{220})+c_{311}S_\mathrm{L}(q,q_{311},\Gamma_{311})+c_{222}S_\mathrm{L}(q,q_{222},\Gamma_{222}) + aq^{-4}$, 
where $c_\mathrm{lq}$, $c_{111}$, $c_{200}$, $c_{220}$, $c_{311}$, and $c_{222}$ are numbers that in principle depend on the Bragg peak multiplicity and Debye-Waller factors but are treated here as fit parameters as is $a$. We note that the crystal structure does not necessarily need to be known in advance in order to use this model. Lorentzian or other suitable line shapes can be used to fit the observable Bragg peaks wherever they occur in a SAXS pattern. If knowledge of the crystal structure is desirable (for example to determine lattice spacings), then it can usually be determined from the number and location of Bragg peaks in a SAXS pattern.\cite{warren_x-ray_1990} Our final fit function is 
$I(q, r,\sigma,c_1,c_2, 
c_\mathrm{lq},q_\mathrm{lq},\Gamma_\mathrm{lq}$,
$c_{111},q_{111},\Gamma_{111}$,
$c_{200},q_{200},\Gamma_{200}$,
$c_{220},q_{220},\Gamma_{220}$,
$c_{311},q_{311},\Gamma_{311}$,
$c_{222},q_{222},\Gamma_{222},a) = I_\mathrm{colloid}(q,r,\sigma,c_1) +  I_\mathrm{SL}(q, r,\sigma,c_2$,
$c_\mathrm{lq},q_\mathrm{lq},\Gamma_\mathrm{lq}$,
$c_{111},q_{111},\Gamma_{111}$,$c_{200},q_{200},\Gamma_{200}$,
$c_{220},q_{220},\Gamma_{220}$,
$c_{311},q_{311},\Gamma_{311}$,
$c_{222},q_{222},\Gamma_{222},a)$. We explicitly fit for $r,\sigma,c_1,c_2$,
$c_\mathrm{lq},q_\mathrm{lq},\Gamma_\mathrm{lq}$,
$c_{111},q_{111},\Gamma_{111}$,$c_{200},q_{200},\Gamma_{200}$,
$c_{220},q_{220},\Gamma_{220}$,
$c_{311},q_{311},\Gamma_{311}$,
$c_{222},q_{222},\Gamma_{222}, \mathrm{and}$ $a$. The fits are done according to a typical least-squares minimization. We seeded our fits by limiting the range of values for most parameters such as the locations of the Bragg peaks based on the locations expected for an \textit{fcc} lattice, the NC size, and the NC size polydispersity to ensure robustness of the fitting procedure. We typically fit the logarithm of the intensities to assist the fitting procedure in capturing the details of smaller features at higher $q$ values. For justification of this model to describe the scattering from finite-sized SLs, see \textbf{Figure S3} and associated text.

\textbf{Power law fits to the SL (111) FWHM.} In order to quantify the coarsening kinetics of the SL (111) FWHM in the experiments, we fit the late-time behavior to a power law, $FWHM \sim t^{-\alpha}$. We fit the data in \textbf{Figure 2e} at late times to focus on the coarsening regime. Each of the power law exponents varies by 10\% at most, depending on the time range over which the data is fit. This uncertainty is smaller than the relative differences between exponents obtained for the different experiments. To facilitate comparison of the different experiments, we chose to fit the data over similar time periods. 

\textbf{Simulations of self-assembly.} Simulations of SL growth and annealing were performed with an underdamped Langevin dynamics in the NVT ensemble in a cubic periodic box using the LAMMPS software\cite{thompson_lammps_2022}. NCs were represented as 10976 spherical particles with pairwise volume exclusion interactions given by a Weeks-Chandler-Andersen (WCA) potential\cite{weeks_role_1971}, \[
    u_\mathrm{WCA}(r)= 
\begin{cases}
    4\epsilon_\mathrm{LJ}[(\frac{\sigma}{r})^{12}-(\frac{\sigma}{r})^6]+\epsilon_\mathrm{LJ},& \text{if } r\le2^\frac{1}{6}\sigma\\
    0,              & \text{otherwise}
\end{cases}
\]
where $\sigma$ is the diameter of the NC, $r$ is the center-to-center distance, and $\epsilon_\mathrm{LJ}$ is the strength of the repulsion. Additionally, NCs interact also \textit{via} a pairwise attractive short-range Morse potential, $u_\mathrm{M}(r)=u_0(e^{-2\alpha(r-r_0)}-2e^{-\alpha(r-r_0)})$ where $r_0=1.23\sigma$ is the position of the minimum of the potential well, $\alpha=10/\sigma$ is the inverse range, and $u_0$ denotes the well depth. The Morse potential has been truncated and shifted to zero at $r=3\sigma$. At a given temperature $T$, the diffusive timescale for NCs is given by  $\tau=\gamma \sigma^2/k_\mathrm{B}T$ where $\gamma$ is the friction coefficient in Langevin dynamics, and $k_\mathrm{B}$ is Boltzmann's constant. We have chosen the density of NCs in the simulation box, mass of NCs, temperature and simulation timestep as $0.01/\sigma^3$, $\gamma^2\sigma^2/\epsilon_\mathrm{LJ}$, $\epsilon_\mathrm{LJ}/k_\mathrm{B}$ and $5\times10^{-5}\tau$, respectively. To compare to experimental timescales, we assume that the NCs follow Stokes’ law of diffusion, with friction coefficient relating to the solvent viscosity $\eta$ as $\gamma=3\pi \eta \sigma$. We then use $\sigma = 4.5$ nm, $\eta= 0.876 \times 10^{-3}$ Pa-s, and $T = 300$ K to obtain $\tau=0.18$ $\mu$s.

NCs were initially equilibrated in the gas phase at $u_0=k_\mathrm{B}T$, before being adiabatically quenched to  $u_0=2.5k_\mathrm{B}T$ and $u_0=2.8 k_\mathrm{B}T$ for the shallow and deeper quench respectively. In the case of the shallow quench, nucleation at this well-depth is prohibitively slow. We thus used a spherical defect-free \textit{fcc} crystal of size 200 NCs as a seed to start crystal growth and annealing. The seed was introduced by excavating a vacuum bubble of radius $1\sigma$ larger than that of the seed in the gas phase simulation, and randomly eliminating gas phase particles to conserve the total particle number. A seed of 100 NCs or smaller disperses into the gas phase. A seed of 200 NCs continues accruing particles till macroscopic size without nucleation of any second cluster in the box throughout the simulation. A seed of 500 NCs gives qualitatively similar results. In the case of the deeper quench, multiple metastable liquid droplets spontaneously nucleate and quickly freeze into crystals. The use of a seed was found irrelevant to the spontaneous nucleation of other clusters, so our reported results for the deeper quench simulation were not obtained with a seed.

\textbf{Crystallinity calculation.} We analyze simulated self-assembly trajectories by tracking crystalline order with Steinhardt-Nelson order parameters\cite{steinhardt_bond-orientational_1983}. At each time frame, we define nearest neighbors to each NCs with a cutoff on the center-to-center distances, $| \mathbf{r}_i-\mathbf{r}_j | \le 1.5\sigma$. We then compute the local bond-orientational order as 
$$\psi_{6m}^{(i)}=\frac{1}{n_i}\sum_{j\in n_i}\bigg( \frac{ \sum_{m=-6}^{6}q_{6m}^{(i)}q_{6m}^{(j)*} }{\sqrt{\sum_{m=-6}^{6}q_{6m}^{(i)}q_{6m}^{(i)*} }\sqrt{\sum_{m=-6}^{6}q_{6m}^{(j)}q_{6m}^{(j)*} }} \bigg)$$
where $n_i$ represent neighbors of the $i$-th particle, and
$q_{6m}^{(i)}=\frac{1}{n_i}\sum_{j\in n_i}Y_{6m}(\Phi_{ij},\theta_{ij})$
with $Y_{lm}(\Phi_{ij},\theta_{ij})$ as the $lm$- spherical Harmonic functions of the angular coordinates of the vector $\mathbf{r}_i-\mathbf{r}_j$. A particle was defined to be locally crystalline if either its $\psi_6^{(i)}$ parameter was above a cutoff of 0.7, or if its neighbor was locally crystalline. All locally crystalline particles are then classified into clusters of direct or indirect neighbors, and particles in clusters of size larger than a cutoff of 100 NCs are defined to be crystals. This cutoff serves to eliminate transient associations of gas-phase NCs that are still less than a critical nucleus size. All reported results about the crystallinity of NCs in the SL phase are computed only over NCs in these large dense locally ordered clusters.

\begin{figure}[ht!]
\centering
\includegraphics[width=14cm]{SI Figures/S1.png}
\caption{Experimental setup at SSRL beamline 1-5. (a) Zoomed out view of the apparatus at the beamline. X-rays are incident from the left. Scattered X-rays enter the SAXS flight tube and scatter onto the CCD detector on the right. (b) Close up picture of the apparatus at the beamline. (c) Face on view of the cuvette and stirrer. The circular recess in the cuvette indicates the location of the 200 $\mu$m window on the front face; a similar recess is found on the opposite side. The blue arrow indicates the direction of rotation of the magnetic stir bar in the cuvette.}
\end{figure}

\begin{figure}[ht!]
\centering
\includegraphics[width=14cm]{SI Figures/S2.png}
\caption{Background subtraction of SAXS patterns. (a) Single SAXS pattern taken from an \textit{in situ} experiment (black dots) and from a background consisting of a cuvette filled with 1.1 M ionic strength salt in hydrazine (red). (b) SAXS pattern obtained after subtraction of the red curve from the black curve in (a). All SAXS patterns used in subsequent analysis were background subtracted in this way.}
\end{figure}

\textbf{Validity of model of SL structure factor.} To test our model for the SL structure factor, we simulated a finite-sized \textit{fcc} SL composed of 10976 NCs 4.5 nm in diameter and computed the structure factor, $S(q)$ (\textbf{Figure S3}). In an experimental scattering configuration, this is the structure factor that would be obtained from an isotropically oriented ensemble of such SLs. The structure factor has two main regimes in $q$: the Bragg scattering at high $q$ and the scattering due to the finite size of the SL at low $q$. In principle, the scattering at low $q$ from this single SL has oscillations with a $\sim q^{-4}$ envelope. These oscillations are difficult to see in \textbf{Figure S3} due to the limited $q$-resolution with which the structure factor was calculated in this region. These oscillations are not present in the experimental SAXS patterns despite them having greater $q$-resolution. The oscillations arise due to the finite size of the SLs, and their exact frequency and phase depend on the size of the SL. When there are many SLs of different sizes as there are in experiment, these oscillations wash out, leaving behind a smooth $\sim q^{-4}$ dependence, which is what is seen in the experimental SAXS patterns. A fit of our model to the structure factor of the simulated finite-sized \textit{fcc} SL is shown in \textbf{Figure S3}. We note that we cut off the low-$q$ term, $aq^{-x}$, at a $q$ value before the appearance of the SL Bragg peaks. In the fitting shown here and in the main text, we chose $q = .1$ $\mathrm{\AA}^{-1}$, but the results of the fitting procedure are insensitive to the exact choice of this cutoff. 

\begin{figure}[ht!]
\centering
\includegraphics[width=10cm]{SI Figures/S3.png}
\caption{Structure factor and fit to structure factor of a simulated finite-sized \textit{fcc} SL. Inset: real space structure of the \textit{fcc} SL from which the structure factor was calculated.}
\end{figure}

\begin{figure}[ht!]
\centering
\includegraphics[width=8cm]{SI Figures/Sresid.png}
\caption{Residuals of a quantitative fit of model to a SAXS pattern. The residuals (data - fit) correspond to the fit to the SAXS pattern shown above. The residuals are normalized to the maximum intensity, $|I|$, of the SAXS pattern that was fit. The magnitude of the residuals is on the order of a few percent at worst and these deviations are largely centered around zero. The structure at low-$q$ results from imperfect background subtraction (see form of background scattering in \textbf{Figure S2}). The discrepancy around the Bragg peak location occurs due to the peak not having a perfectly Lorentzian line shape. These slight discrepancies between the fitted model and the data do not limit the utility of the fit in quantifying the amount of NCs in the colloidal phase and the location and width of the (111) Bragg peak. 
}
\end{figure}

\begin{figure}[ht!]
\centering
\includegraphics[width=10cm]{SI Figures/S4v2.png}
\caption{Close-up of low density binodal of the phase diagram extracted from experiments. Error bars are derived as in \textbf{Figure 2c} of the main text. The open circle indicates a ($\phi,\lambda$) phase diagram location in which no SL formation was observed. The black phase boundary curve is drawn in as a visual guide. The open circle bounds the height of the phase boundary curve at $\phi \sim 0.0024$.}
\end{figure}

\begin{figure}[ht!]
\centering
\includegraphics[width=14cm]{SI Figures/S5.png}
\caption{Structure factors and fits of the particles in the simulations. (a) Shallow quench structure factor and fit. (b) Deeper quench structure factor and fit.}
\end{figure}

\textbf{Volume fraction of the SL phase as a function of quench depth.} To understand why the volume fraction of the SL phase increases as the quench depth increases, we first compute the nearest neighbor distance between NCs in the SL phase with the following procedure. From fitting the SAXS patterns of the colloidal NCs (\textbf{Figure 2a} main text), we extract a Gaussian NC size distribution of 4.5 ± 0.3 nm (\textbf{Figure S7a}). From the fits to the SL \textit{fcc} (111) Bragg peaks in the experimental and simulated SAXS patterns, we extract the SL lattice constants using $a = 2\pi\sqrt{3}/q_{111}$. We convert the lattice constants into center-to-center nearest neighbor distances by dividing a by 2 (\textbf{Figure S7b, S7c}). Using the extracted NC size distribution, we compute the surface-to-surface nearest neighbor distance as a function of time and quench depth (\textbf{Figure S7b, S7c}). We find that as a function of quench depth, the NCs in the deepest experimental quench are on average $\sim$2.8 $\mathrm{\AA}$ closer together than the NCs in the shallowest experimental quench. One possible explanation for this trend is that at deeper quenches, the interactions are stronger and the particles are on average closer together. In the simulations, the NCs in the SLs in the deeper quench simulation are on average $\sim$ 0.8 $\mathrm{\AA}$ closer together than the NCs in the SL in the shallow quench simulation. 

Another possible explanation is that the NCs selectively quench depending on their size. In other words, the largest NCs, which experience the greatest van der Waals attraction, might incorporate into the SL phase before the smallest NCs do. To test this hypothesis, we performed a simple numerical test. First, we calculated the magnitude of the dispersion force between NCs as a function of their radii in \textbf{Figure S7d}. We find the dispersion force can vary by up to 1 $k_\mathrm{B}T$ at room temperature over the NC sizes present here. This variation can be substantial compared to the overall attractive force between NCs, which is $\sim$ several $k_\mathrm{B}T$. 

\begin{figure}[ht!]
\centering
\includegraphics[width=14cm]{SI Figures/S6.png}
\caption{Size selectivity in NC SL self-assembly. (a) NC size probability distribution obtained from SAXS fitting. (b) center-to-center and surface-to-surface nearest neighbor distances between NCs in the SL phase as a function of time and quench depth in experiments. Color scheme is the same as \textbf{Figure 2} in the main text. (c) same as (b) except for NCs in the simulated SLs. (d) van der Waals force between NCs at a surface-to-surface separation of 1 nm as a function of NC diameter.}
\end{figure}

Second, to further test this hypothesis, we calculated the expected difference in nearest neighbor distances assuming that size selectivity occurs. At the shallowest quench depth, 60\% of the original colloidal NCs end up in the SL phase at the end of our measurement, while at the deepest quench depth, 98\% of the original colloidal NCs end up in the SL phase. If we assume the largest 60\% and largest 98\% of NCs selectively form SLs in each of those quenches, the average size of the 98\% subpopulation is $\sim$2.1 $\mathrm{\AA}$ smaller than that of the 60\% subpopulation. The magnitude and direction of this difference is consistent with the $\sim$2.8 $\mathrm{\AA}$ decrease in nearest neighbor distance found in \textbf{Figure S7b} when comparing the same respective quench depths. In conclusion, this polydispersity could account for a large portion of the difference in nearest neighbor distances between the shallowest and deepest quenches. We note that the SLs contract as a function of time across quench depths, decreasing by $\sim$0.2 - 0.4 $\mathrm{\AA}$. These are very small changes that could be due to rearrangements of overlapping ligand shells on neighboring NCs.

\textbf{Equilibrium NC fractions in colloidal and SL phases.} We calculated the equilibrium volume fraction of the colloidal phase in \textbf{Figure 2c} in the main text based on the fraction of NCs in the colloidal phase $\sim$1 hour post-quench. The colloidal NC fractions decrease by $\sim$0.01-1\% over a one hour time window starting after the initial quenches. Therefore, we estimate that the colloidal NC fractions we obtained $\sim$1 hour post-quench should be within a few percent of the true equilibrium colloidal NC fractions. We calculated the volume fraction of the SL phase based on the $q$ value on which the SL \textit{fcc} (111) peak is centered for SLs $\sim$1 hour post-quench. The SLs $\sim$1 hour post-quench have not reached equilibrium. Deviations from equilibrium include the finite-size of the SLs and any strain/disorder in the SLs. These deviations from equilibrium primarily contribute to the FWHM of the Bragg peaks and not the locations of the Bragg peaks. As a result, we expect that the volume fraction of the SL phase that we extracted to be a good estimate of the equilibrium SL volume fraction. One exception to this hypothesis is isotropic strain: SLs with isotropic strain have Bragg peaks shifted to lower $q$ values than those of unstrained SLs and higher order Bragg peaks are more affected than lower order Bragg peaks\cite{khorsand_zak_x-ray_2011}. The SLs formed in this work have Bragg peaks that are well indexed to those of an \textit{fcc} crystal, indicating that any strain in the SLs is likely not isotropic in nature. Another possible exception is as follows: we might naively expect that for large amounts of disorder, the density of the condensed phase would decrease as close packing is no longer achieved, \textit{i.e.}, a lower-density gel or aggregate forms instead of a SL. We are not in this regime for any of the experiments performed in this work. The more disordered SLs formed at deeper quenches, in fact, have \textit{fcc} (111) peaks located at higher $q$ values (more dense) than more ordered SLs formed at shallower quenches.

\begin{figure}[ht!]
\centering
\includegraphics[width=12cm]{SI Figures/S7.png}
\caption{Short-lived or absent metastable liquid phase in \textit{in situ}  SAXS data. (a) Number of NCs in dense (black points) and crystalline (red points) phases for the largest cluster in the deeper quench simulation. See \textbf{Methods} section of the main text for definition of $\tau$. SL nucleation occurs between 500-1000$\tau$ or 100-200 $\mu$s. (b) Schematic phase diagram of particles interacting with short-range interactions. The colloid-metastable liquid binodal drawn here peaks at lower $k_\mathrm{B}T/u_0$ than the binodal in \textbf{Figure 2b}. The brown arrows indicate schematic quenches from the colloid to colloid-SL coexistence regions.}
\end{figure}

\begin{figure}[ht!]
\centering
\includegraphics[width=14cm]{SI Figures/S8v2.png}
\caption{Interparticle potentials for shallow, intermediate, and deep quenches. Interparticle potential, $u(r)$, between two NCs as a function of their center-to-center distance parameterized relative to the effective size of the NCs (left) and in nm (right). Shallow quenches are shown in green, intermediate quenches are shown in gold, and deep quenches are shown in red. The depths of the potential were chosen to span the $\sim$2.5 - 6 $k_\mathrm{B}T$ range inferred in the main text. The range of the potentials was approximated to be $\sim$20\% of the effective NC size for each quench depth. We then assume the effective size of the NCs corresponds to the center-to-center distance between NCs in the SL phase in \textbf{Figure S7b}. We multiply the center-to-center distance in $r/\sigma$ by the effective size of the NCs to obtain the range in nm.}
\end{figure}

\FloatBarrier
\bibliography{main.bib}